\def\dir{./}
\documentclass[useAMS,usenatbib]{\dir mn2e}
\usepackage[fleqn]{amsmath}
\usepackage{amssymb}
\usepackage[super]{nth}
\usepackage{natbib}
\usepackage{graphicx}
\usepackage{float}
\usepackage{mathrsfs}
\usepackage{mathtools}
\usepackage{multirow}
\usepackage[usenames,dvipsnames]{color}
\usepackage{\dir aas_macros}
\usepackage{comment}
\usepackage{subfigure}
\usepackage{hyperref}
\usepackage{mathrsfs}
\usepackage{epstopdf, dcolumn}
\usepackage{wasysym}
\usepackage{dashrule}
\usepackage{xcolor}
\usepackage{textcomp}
\usepackage{threeparttable}

\definecolor{color_meraxes_line_1}{RGB}{27,158,119}
\definecolor{color_meraxes_line_2}{RGB}{217,95,2}
\definecolor{color_smaug_line_1}{RGB}{117,112,179}
\definecolor{color_smaug_line_2}{RGB}{231,41,138}

\definecolor{color_meraxes_fiducial}{RGB}{230,171,2}
\definecolor{color_meraxes_S}{RGB}{102,166,30}
\definecolor{color_meraxes_SM}{RGB}{231,41,138}
\definecolor{color_meraxes_SMB}{RGB}{217,95,2}
\definecolor{color_meraxes}{RGB}{27,158,119}

\newcommand{\appropto}{\mathrel{\vcenter{
 \offinterlineskip\halign{\hfil$##$\cr
 \propto\cr\noalign{\kern2pt}\sim\cr\noalign{\kern-2pt}}}}}
\newcommand{\Rom}[1]{\uppercase\expandafter{\romannumeral #1}}
\newcommand{\rom}[1]{\lowercase\expandafter{\romannumeral #1}}
\newcommand{\meraxes}{\textsc{Meraxes}}
\newcommand{\smaug}{{\textit{Smaug}}}
\newcommand{\nbody}{\textit{N}-body}
\newcommand{\htwo}{\mathrm{H}_2}

\defcitealias{Mutch2016a}{Paper-\Rom{3}}
\defcitealias{Qin2017a}{Paper-\Rom{8}}

\title[DRAGONS \Rom{14}: dwarf galaxies]{\LARGE Dark-ages~Reionization~and~Galaxy~Formation~Simulation~-~\Rom{14}.\\ Gas accretion, cooling and star formation in dwarf galaxies at high redshift}

\author[Qin et al.]{Yuxiang Qin$^{1,2}$\thanks{E-mail: Yuxiang.L.Qin@Gmail.com}, Alan R. Duffy$^{3,2}$, Simon J. Mutch$^{1,2}$, Gregory B. Poole$^{1,3}$,
\newauthor Paul M. Geil$^1$, Andrei Mesinger$^4$ and J. Stuart B. Wyithe$^{1,2}$\thanks{E-mail: swyithe@unimelb.edu.au}\\
$^{1}$School of Physics, University of Melbourne, Parkville, VIC 3010, Australia\\
$^{2}$ARC Centre of Excellence for All Sky Astrophysics in 3 Dimensions (ASTRO 3D)\\
$^{3}$Centre for Astrophysics and Supercomputing, Swinburne University of Technology, PO Box 218, Hawthorn VIC 3122, Australia\\
$^{4}$Scuola Normale Superiore, Piazza dei Cavalieri 7, I-56126 Pisa, Italy}
\begin{document}

%\date{Accepted 2014 December 2. Received 2014 December 1}
\date{\today\ draft - \nth{22}}
\pagerange{\pageref{firstpage}--\pageref{lastpage}} \pubyear{2018}
\maketitle
\label{firstpage}

\begin{abstract}
We study dwarf galaxy formation at high redshift ($z\ge5$) using a suite of high-resolution, cosmological hydrodynamic simulations and a semi-analytic model (SAM). We focus on gas accretion, cooling and star formation in this work by isolating the relevant process from reionization and supernova feedback, which will be further discussed in a companion paper. We apply the SAM to halo merger trees constructed from a collisionless \textit{N}-body simulation sharing identical initial conditions to the hydrodynamic suite, and calibrate the free parameters against the stellar mass function predicted by the hydrodynamic simulations at $z=5$. By making comparisons of the star formation history and gas components calculated by the two modelling techniques, we find that semi-analytic prescriptions that are commonly adopted in the literature of low-redshift galaxy formation do not accurately represent dwarf galaxy properties in the hydrodynamic simulation at earlier times. We propose 3 modifications to SAMs that will provide more accurate high-redshift simulations. These include 1) the halo mass and baryon fraction which are overestimated by collisionless \textit{N}-body simulations; 2) the star formation efficiency which follows a different cosmic evolutionary path from the hydrodynamic simulation; and 3) the cooling rate which is not well defined for dwarf galaxies at high redshift. {\color{black}Accurate semi-analytic modelling of dwarf galaxy formation informed by detailed hydrodynamical modelling will facilitate reliable semi-analytic predictions over the large volumes needed for the study of reionization.}
\end{abstract}


\section{Introduction}
About 150 Myr after the Big Bang, the Universe entered the Epoch of Reionization (EoR; \citealt{Planck2016A&A...596A.108P}), when the first stars and galaxies were created and started to photoionize the intergalactic neutral hydrogen. Over the past decade, there have been significant advances in the study of galaxies during the EoR and the evolution of the ionized intergalactic medium (IGM). High-redshift galaxies thought to have sourced reionization have been discovered all the way to $z\sim11$ \citep{Oesch2016}, when the Universe was only 400 Myr old. However, the sample of these early galaxies remains small with only ${\sim}1000$ candidates at $z>6$ identified using advanced space-based instruments \citep{Bouwens2014,Bouwens2015}. Achieving a physical understanding between galaxies and the progress of reionization is therefore aided by numerical simulations (see e.g. \citealt{Bertschinger1998,Baugh2006,Dolag2008,Somerville2015}). 

Two of the main numerical approaches to model galaxy formation are hydrodynamic simulations and semi-analytic models (SAMs) applied to dark matter halo merger trees constructed from cosmological \textit{N}-body simulations. Since they adopt different approximations, their strengths and requirements are diverse. Hydrodynamic simulations include baryons as well as dark matter, and simulate the complex baryonic physics more directly (e.g. \citealt{Crain2009,Schaye2010,vogelsberger2014properties,Schaye2014,feng2015bluetides,Pawlik2017}). However, while galaxies form and evolve in the presence of gravity and hydrodynamical forces in a simulation volume, its large computational expense limits the capability of simultaneously resolving small-scale regions and capturing massive systems. For instance, in order to study reionization and galaxy formation during the EoR, simulations are required to have a volume size of at least $10^6 \mathrm{Mpc}^3$ \citep{Iliev2014MNRAS.439..725I} and a particle resolution that is able to resolve haloes with masses around $5\times10^7\mathrm{M}_\odot$ (i.e. ${>}10$ billion particles; \citealt{Barkana1999ApJ...523...54B}). While it remains challenging to explore the relevant parameter space in hydrodynamic simulations, SAMs, with their advantages of low expense on computational resources, become the alternative in this case.

\subsection{Semi-analytical models}\label{sec:SAMs}
\textit{N}-body simulations (e.g. \citealt{Navarro1997,springel2005simulating,iliev2008simulating,Boylan_Kolchin2009,Klypin2010,Garrison2017}) solve for the gravitational force on each collisionless dark matter particle but neglect the baryonic physics. In order to implement the galaxy formation physics of baryons into these simulations, halo merger trees are constructed and based on the properties inherited from the merger trees, SAMs parametrize baryonic processes to evolve galaxies, offering an efficient approach to simulate galaxies in a cosmological context \citep[e.g.][]{Cole2000,Hatton2003,Baugh2005,croton2006many,DeLucia2007,Somerville2008,guo2011dwarf,Henriques2015}.

In general, a SAM first designates haloes as galaxies and endows them with stellar components and several gas reservoirs of varied functionalities. The latter usually include a cold gas disc where star formation occurs and a hot halo of non-star-forming gas. Mass is then manipulated and transferred between these baryonic sectors as a result of varied baryonic processes including accretion, cooling, star formation and feedback from supernovae or active galactic nuclei (AGN). These processes are implemented using scaling functions of astrophysics that are motivated either from first principles (e.g. the thermal cooling rate, see Section \ref{sec:SAM_cooling}), from observations [e.g. the Kennicutt--Schmidt (KS) star formation law, see Section \ref{sec:SAM_SF}], or from more complicated simulations such as hydrodynamic and radiative transfer calculations (e.g. reionization feedback, \citealt{Sobacchi2013a}). 
%obtains the halo properties, such as virial mass, position and spin from a parent \textit{N}-body simulation with detailed particle distributions of the halo simplified by a universal density profile (e.g. isothermal). 

%Galaxy properties are then calculated 

Depending on the scientific goal and applicable range, the complexity of SAMs ranges from the treatment of halo profiles to the implementation of baryonic physics. Taking star formation as an example, it can be modelled in a simplified manner by consuming the entire cold gas disc within a given time-scale (e.g. {\sc sage}, \citealt{croton2006many}), or as \citealt{Stevens2016} demonstrated (in the updated version, {\sc dark sage}), by
\begin{enumerate}
	\item considering the disc as a combination of a series of gas rings with angular momentum conserved between each annulus and transferred dynamically; 
	\item using the mid-plane pressure inferred from local observations and theoretical work (e.g. \citealt{Blitz2006,Krumholz2009}) to split the disc into molecules and atoms, and forming stars directly from the giant $\htwo$ clouds; and
	\item including disc instabilities and forming additional components representing stellar bulges, which might possess different ages or dynamics from the stellar disc.
\end{enumerate}

Despite these improvements recently implemented into modern SAMs, some prescriptions are still commonly adopted by many of these models. For instance, in order to initialize the baryonic component of haloes from dark matter only \textit{N}-body simulations, the universal baryon fraction is applied to every virialized system with a further suppression due to reionization feedback. These baryons are considered as infalling gas and are usually assumed to share the virial temperature of the host halo as a result of experiencing shock--heating. {\color{black}However, we note that the degeneracy between parameters of SAMs, as well as their time and mass dependencies introduced by these simplified baryonic prescriptions are not well understood. In addition, lack of available observational data means that semi-analytic predictions have generally been tested against observables that are massive and in the nearby Universe \citep{croton2006many,guo2011dwarf,Stevens2016}. Their consistency and performance in the low-mass regime or at high redshift remain unclear -- a model might be able to reproduce a limited number of observed quantities by either careful calibrations or by introducing more free parameters accompanied by \textit{physical motivations}, but incorrectly predict the other properties that cannot be observed at this stage.

This motivates us to 1) test the performance of SAMs at high redshift, which were originally developed from the literature of low-redshift galaxy formation; and 2) inform more accurate semi-analytic modelling of dwarf galaxies, which will be applied to large-volume simulations to study the EoR in the future. In the absence of detailed observational data, constraints on the properties of SAM galaxies can be supplemented through comparisons against physics-rich hydrodynamic simulations, which are expected to predict galaxies that are more representative than simplified calculations using semi-analytic approaches (e.g. \citealt{Guo2016,mitchell:2017je,stevens:2017fi,cote:2017uh}). In this work, we take the {\meraxes} SAM \citep{Mutch2016a} as an example, and apply it to the halo merger trees constructed from a collisionless \textit{N}-body simulation sharing identical initial conditions to the {\smaug} suite of hydrodynamic simulations \citep{duffy2014low}. We calibrate the SAM against the {\smaug} hydrodynamic simulations and make direct comparisons of the calculated galaxy properties using the two modelling techniques. Since the hydrodynamic simulations were run with different assumptions of galactic or stellar feedback, the free parameters of the SAM are chosen accordingly to reproduce the predicted galaxy properties from the corresponding hydrodynamic simulation.}
	
This paper focuses on modelling of gas accretion, cooling and star formation, the implementation of which is common among many modern SAMs \citep{croton2006many,Somerville2008,guo2011dwarf,Henriques2015}. The effect of feedback on star formation (which varies between models) is excluded and will be presented in a companion paper (Qin et al. in prep.). We begin with a brief introduction of the \nbody/hydrodynamic simulations and SAM utilized in this work in Section \ref{sec:models}. Then we discuss the consequence of applying SAMs directly to collisionless \textit{N}-body simulations and propose modifications accordingly in Section \ref{sec:halo mass}. In Section \ref{sec:results}, we proceed with the proposed modifications and calibrate the SAM to reproduce the hydrodynamic result. We investigate galaxy properties in detail and make comparisons between the hydrodynamic simulation and the SAM results. Based on these, we propose additional modifications to the SAM to inform better modelling of dwarf galaxies at high redshift. We conclude in Section \ref{sec:conclusion}. In this work, we adopt cosmological parameters from \textit{WMAP7} ($\Omega_{\mathrm{m}}, \Omega_{\mathrm{b}}, \Omega_{\mathrm{\Lambda}}, h, \sigma_8, n_s $ = 0.275, 0.0458, 0.725, 0.702, 0.816, 0.968; \citealt{Komatsu2011}) in all simulations.

\section{The Dragons Project}\label{sec:models}
The Dark-ages Reionization And Galaxy Formation Observable from Numerical Simulation (DRAGONS\footnote{\href{http://dragons.ph.unimelb.edu.au/}{http://dragons.ph.unimelb.edu.au}}) project employs \textit{N}-body simulations \citep{Poole2016}, hydrodynamic simulations \citep{duffy2014low,Qin2017a} and SAMs \citep{Mutch2016a,Qin2017c} to study reionization and galaxy formation at high redshift ($z \ge 5$, \citealt{Geil2016,Geil2017,Liu2016,Liu2017,Mutch2016b,Park2017,Duffy2017,Qin2017b}). The resulting galaxy catalogues from both the hydrodynamic simulation and SAM with feedback included are in good agreement with observations including the stellar mass and galaxy UV luminosity functions across cosmic time.

In this work, we focus on high-redshift modelling ($z\ge5$) of currently unobservable dwarf galaxies ($M_\mathrm{vir}\lesssim10^{10.5}\mathrm{M}_\odot$) which are believed to be the dominant source of ionizing photons driving reionization \citep{duffy2014low,Liu2016}. Using simplified models, we isolate prescriptions of gas accretion, cooling and star formation from reionization and supernova feedback\footnote{AGN feedback, which is expected to have an insignificant impact to the formation of dwarf galaxies at high redshift, is not considered here.}, which will be further discussed in a companion paper. In the following two subsections, we summarize the DRAGONS SAM and hydrodynamic simulation, named {\meraxes} and {\smaug}, respectively.

\subsection{\textsc{Meraxes}}\label{sec:meraxes}
{\meraxes} is based on the \textit{Munich} SAM \citep{croton2006many,guo2011dwarf} and optimized to study galaxy formation at high redshift. Based on halo properties read from merger trees, it evolves galaxies using simplified prescriptions, which calculate baryonic infall, cooling, star formation, supernova feedback, metal enrichment, stellar mass recycling, AGN feedback and mergers. In addition, {\meraxes} incorporates 21cm{\sc fast} \citep{Mesinger2010}, a semi-numerical approach to evolve ionization fields, for the purpose of investigating reionization feedback and exploring the IGM state during the EoR. We refer the interested reader to \citet[hereafter \citetalias{Mutch2016a}]{Mutch2016a} and \citet{Qin2017c} for a detailed description of {\meraxes} and provide a brief review of the relevant modelling prescriptions in this section.

\subsubsection{Accretion (and stripping)}\label{sec:SAM_accretion}
To connect \textit{N}-body simulations of dark matter with baryonic physics, the first implementation is the accretion of gas. Following convention, {\color{black}a universal baryon fraction ($f_\mathrm{b}=\Omega_\mathrm{b}/\Omega_\mathrm{m}$) with suppression due to reionization heating ($f_\mathrm{mod}$) is applied and each halo group (i.e. {\sc fof} group, see Appendix \ref{sec:dark matter halo merger trees}), with a total mass of $M_\mathrm{vir}$, accretes gas from the IGM}
\begin{equation}\label{eq:accretion}
\Delta m_\mathrm{hot} = f_\mathrm{mod}f_\mathrm{b}M_\mathrm{vir}-\sum\limits_{N_\mathrm{halo}} m_\mathrm{baryon},
\end{equation}
where $m_\mathrm{baryon}$ is the total baryonic mass including\footnote{Gas that has been ejected from galaxies due to supernova feedback is not included in this work. We refer the interested readers to \citetalias{Mutch2016a} for more details about feedback and gas stripping prescriptions.} hot gas ($m_\mathrm{hot}$), cold gas ($m_\mathrm{cold}$) and stellar mass ($m_*$), and the summation symbol indicates that baryons of all $N_\mathrm{halo}$ haloes including the central and satellite haloes are considered. $f_\mathrm{mod}\le1$ represents a baryon fraction modifier accounting for the suppression due to the reionization heating background. 

In the case of $\Delta m_\mathrm{hot}>0$, all of the mass is assumed to be accreted by the central halo due to its dominant gravitational potential in the entire virialized system, and are stored in its hot gas reservoir. However, when $\Delta m_\mathrm{hot}<0$, baryons will be removed from the hot gas reservoir of the central halo in order to maintain the universal baryon fraction. Note that baryons that have reached the cold gas disc or have formed stars are assumed to be gravitationally bound, and do not get further removed even if $f_\mathrm{mod}f_\mathrm{b}M_\mathrm{vir} - \sum\limits_{N_\mathrm{halo}}\left(m_*+m_\mathrm{cold}\right)<0$.

\subsubsection{Cooling}\label{sec:SAM_cooling}
Gas falling into a halo is stored in the hot gas reservoir, which is assumed to share the halo virial temperature, $T_\mathrm{vir}$, due to shock heating. It then cools within the following time-scale
\begin{equation}\label{eq:t_cool}
t_{\mathrm{cool}}\left(r\right) = \dfrac{3\bar{\mu}m_\mathrm{p}kT_\mathrm{vir}}{2\rho_\mathrm{hot}\left(r\right)\Lambda\left(T_\mathrm{vir}\right)},
\end{equation}
where $\bar{\mu}=0.59$, $m_\mathrm{p}$, $k$, $\rho_\mathrm{hot}$ and $\Lambda$ are the mean molecular weight of fully ionized gas, the mass of a proton, the Boltzmann constant, the hot gas density and the cooling function \citep{Sutherland1993} that is determined by the temperature\footnote{{\color{black}Metals are predominately created by massive stars that have reached type \Rom{2} supernovae. With supernova feedback included, metals are released into the cold gas disc, which are further transferred between various gas reservoirs through heating, cooling and reincorporation. However, these are not considered in this work and $\Lambda$ only accounts for primordial elements.}}. Assuming the hot gas follows a singular isothermal sphere (SIS) profile, we calculate the cooling radius, $r_\mathrm{cool}$, at which the cooling time is equal to the halo dynamical time, $t_\mathrm{dyn}$, through
\begin{equation}
r_\mathrm{cool} = \sqrt{\dfrac{m_\mathrm{hot}\Lambda\left(T_\mathrm{vir}\right)}{6\pi \bar{\mu}m_\mathrm{p}kT_\mathrm{vir}V_\mathrm{vir}}},
\end{equation}
where $V_\mathrm{vir}$ is the virial velocity of the host halo. 

Based on the ratio of the cooling radius to the virial radius, $R_\mathrm{vir}$, we calculate the cooling rate and consider the following two scenarios:
\begin{enumerate}
	\item $r_\mathrm{cool}<R_\mathrm{vir}$ (\textit{static hot halo}): cooling is assumed in thermal equilibrium and the cooling rate is determined by the continuity equation;
	\item $r_\mathrm{cool}\geq R_\mathrm{vir}$ (\textit{rapid cooling}): cooling is treated as free-fall collapse and all of the hot gas component becomes cold within the dynamical time-scale, $t_\mathrm{dyn}$.
\end{enumerate}
Therefore, the cooling rate can be calculated by
\begin{equation}\label{eq:mcool}
\dot{m}_{\mathrm{cool}}{=}\dfrac{m_{\mathrm{hot}}}{t_\mathrm{dyn}}\times\min\left(1,\dfrac{r_{\mathrm{cool}}}{R_{\mathrm{vir}}}\right).
\end{equation}

\subsubsection{Star formation}\label{sec:SAM_SF}
Cooling leads to the build-up of a cold gas reservoir, which is assumed to form a rotationally supported disc, where stars form, with an exponential surface density profile. When the cold gas mass, $m_\mathrm{cold}$, exceeds a critical value, $m_\mathrm{crit}$, we use a phenomenological relation suggested by observations \citep{kennicutt1998global} to calculate the star formation rate (SFR)\footnote{The second channel of star formation is through mergers, which does not have a significant impact in our model.}
\begin{equation}\label{eq:sfr} 
\dot{m}_{\star} {=} \alpha_{\mathrm{sf}}\times\dfrac{\mathrm{max}\left(0,m_{\mathrm{cold}}{-}m_{\mathrm{crit}}\right)}{t_{\mathrm{dyn,disc}}},
\end{equation}
where $t_{\mathrm{dyn,disc}}\equiv1.5\sqrt{2}\lambda t_{\mathrm{dyn}}$ is the dynamical time of the cold gas disc inferred by the halo spin parameter {\color{black}\citep{bullock2001apj...555..240b}}, $\lambda$, with an additional assumption that the specific angular momentum is conserved between the cold gas disc and the host halo \citep{Mo1998}. A free parameter, $\alpha_{\mathrm{sf}}$, is introduced for the purpose of adjusting the star formation efficiency.

We note that, in order to be consistent with the hydrodynamic simulations introduced in the next subsection, the newly formed stars are assumed to follow an initial mass function in the mass range of $0.1-120\mathrm{M}_\odot$ with the following form \citep{chabrier2003galactic}
\begin{equation}\label{eq:imf}
\phi(m) = 
\begin{cases}
0.85m^{-1} \mathrm{exp}\left[-\dfrac{\left(\mathrm{log_{10}}m - \mathrm{log_{10}} 0.079\right)^2}{2\times0.69^2}\right],\\
\hfill \text{if}\ 0.1\mathrm{M}_\odot \leq m < 1\mathrm{M}_\odot;\\
\\
0.24 m^{-2.3},\hfill \text{if}\ 1 \leq m \leq 120\mathrm{M}_\odot. 
\end{cases}
\end{equation}

\subsection{{\smaug}}\label{sec:smaug}
{\smaug} is a series of high-resolution hydrodynamic simulations (\citealt{duffy2014low}, \citealt{Qin2017a}) which were run with a modified {\sc gadget}-2 \textit{N}-body/hydrodynamics code \citep{springel2005cosmological}. These models follow the OverWhelmingly Large Simulations project (OWLS; \citealt{Schaye2010}), in which the full hydrodynamic simulation that involves star formation \citep{Schaye2008}, radiative cooling \citep{Wiersma2009a}, supernova feedback \citep{DallaVecchia2008, DallaVecchia2012} and reionization feedback \citep{Haardt2001}, has been shown to successfully reproduce observed galaxy properties at low redshift including the cosmic star formation history \citep{Schaye2010} and the stellar mass function at $z<2$ \citep{Haas2013}. In addition, \citet{Katsianis2017} recently showed that the EAGLE simulations, an extension of OWLS, are also in agreement with the observed SFR function across cosmic time ($z\sim 0-8$). These achievements motivate us to make use of the hydrodynamic simulations and to investigate the validity of semi-analytic prescriptions at high redshift where observational constraints are unavailable.

\subsubsection{Galaxy physics}\label{sec:galaxy physics}
In this work, the simulation from {\smaug} comprises $512^3$ baryon and $512^3$ dark matter particles within a cube of comoving side of $10 h^{-1}\mathrm{Mpc}$. This equates to a mass resolution of $4.7 (0.9){\times} 10^{5} h^{-1}\mathrm{M}_\odot$ per dark matter (gas) particle. The time interval between two outputs is around 11 Myr, and the initial conditions were generated with the {\sc grafic} package \citep{Bertschinger2001} at $z=199$ using the Zel'dovich approximation \citep{Zeldovich1970}. We summarize the subgrid physics adopted in {\smaug} in terms of cooling and star formation for comparison with the semi-analytic prescriptions (see Section \ref{sec:meraxes}), and refer the interested reader to \citet{Schaye2010} and \citet{duffy2014low} for more details.
\begin{enumerate}
	\item \textit{Cooling} \citep{Wiersma2009a} consists of only primordial elements and is pre-tabulated using the {\sc cloudy} package \citep{ferland1998cloudy}. The cooling rate is calculated in the presence of the cosmic microwave background (CMB), accounting for both free--free scattering between gas particles and Compton scattering between gas particles and CMB photons.

	\item \textit{Star formation} \citep{Schaye2008} occurs by stochastically converting gas particles to star particles in the ISM. In practice, the local density around a gas particle, $\rho_\mathrm{g}$, can be inferred by smoothing the particle mass, $m_\mathrm{g}$, within a certain volume. When $\rho_\mathrm{g}$ exceeds a critical value, the gas particle is assumed to be multiphase ISM state with pressure $P\propto \rho^{\gamma_\mathrm{eff}}$, where $\gamma_\mathrm{eff}=\dfrac{4}{3}$ represents the effective ratio of specific heats. Note that resolving star forming regions in a cosmological hydrodynamic simulation is computationally unrealistic. Therefore, the gas depletion time, $t_\mathrm{g}$, in the hydrodynamic simulation is calculated based on the observed KS (\citealt{kennicutt1998global}) star formation law, and the SFR is given by $\dot{m}_* = m_\mathrm{g}/t_\mathrm{g}$. We note that although the star formation recipes adopted in the SAM and hydrodynamic simulation are both developed from the KS law, there are discrepancies between the two numerical approaches, which are further discussed in Section \ref{sec:sfefficiency}.
\end{enumerate}

\subsubsection{Simulations}\label{sec:simulations}
The two simulations used in this work are
\begin{enumerate}
	\item[(1)] \textit{DMONLY}, a collisionless \textit{N}-body simulation performed using the same initial conditions from the full simulation but neglecting hydrodynamic forces from the baryonic component. It consists of $512^3$ collisionless particles with masses of $5.7\times10^5h^{-1}\mathrm{M}_\odot$ within a cube of comoving side of $10h^{-1}\mathrm{Mpc}$. In order to apply the {\meraxes} SAM and compare the result with {\smaug}, \textit{DMONLY} is used to construct dark matter halo merger trees (See Appendix \ref{sec:match});
	\item[(2)] \textit{NOSN\_NOZCOOL\_NoRe}, a toy model with cooling in the absence of metal line emissions and reionization heating. It ignores feedback from exploding supernovae and heating due to reionization, and will be used to compare with the SAM for the purpose of investigating gas accretion, cooling and star formation without impacts from any feedback. The top left panel of Fig. \ref{fig:indicator} presents an example of the gas density distribution at $z=5$ in the \textit{NOSN\_NOZCOOL\_NoRe} {\smaug} simulation.	
\end{enumerate} 

\subsubsection{Cold gas (star-forming gas)}\label{sec:cold gas}
In the SAM, cold gas is considered as potentially star-formation gas (see Section \ref{sec:SAM_SF}). In order to facilitate direct comparisons of the gas reservoir with the semi-analytic calculation, we specify, for each galaxy in the hydrodynamic simulation, a star-formation gas component. This reservoir comprises all potential star forming particles of the galaxy, making it comparable to the assumption adopted in the SAM.

As mentioned before, a gas particle located in a cold dense region is considered as multiphase and can potentially form stars. More specifically in {\smaug}, this requires the particle to satisfy \textit{A} AND \textit{B} AND (\textit{C} OR \textit{D}), a combination of the following four criteria \citep{Schaye2010}:
\begin{enumerate}
	\item[\textit{A}.] a high physical density, the threshold of which is a fixed value ($0.1\mathrm{cm}^{-3}$) and informs the formation of a multiphase ISM through gravitational instability;
	\item[\textit{B}.] a high comoving density, the threshold of which is 57.7 times the cosmic mean and prevents spurious star formation in less dense region at high redshift;
	\item[\textit{C}.] a high physical density and a low temperature, the thresholds of which are informed from the equation of state of the unresolved warm dense ISM, and can recover the KS star formation law;
	\item[\textit{D}.] a low temperature, the threshold of which is a constant ($10^{5}\mathrm{K}$) and is proposed to capture cold ISM in less dense regions.
\end{enumerate}

The bottom panels of Fig. \ref{fig:temp_nosn_nozcool} show the gas density--temperature phase diagram of two haloes at $z=5$ in the \textit{NOSN\_NOZCOOL\_NoRe} {\smaug} simulation for illustration. Particles located in the lower left region are identified as star-formation gas; otherwise they are non-star-forming gas. 

\subsubsection{Hot gas}\label{sec:hot gas}
\begin{figure}
	\begin{minipage}{\columnwidth}
		\centering
		\includegraphics[width=\textwidth]{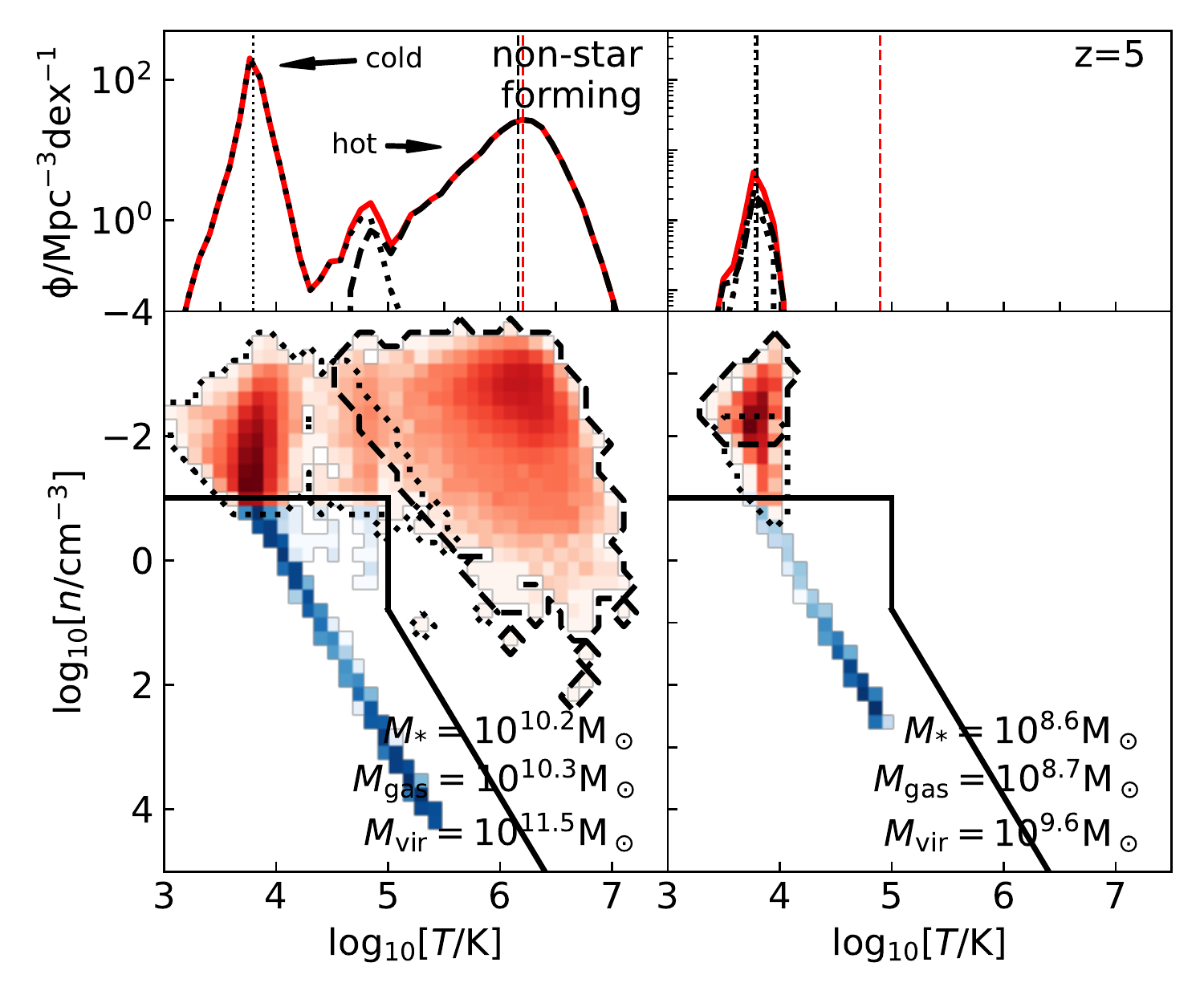}
	\end{minipage}
	\caption{\label{fig:temp_nosn_nozcool}\textit{Bottom panels:} gas density--temperature phase diagram of two haloes at $z=5$ in the \textit{NOSN\_NOZCOOL\_NoRe} {\smaug} simulation. The stellar mass, gas mass and halo mass are shown on the lower right corner. The piecewise function shown with black solid lines indicates the separation of gas particles with different phases -- particles inside (outside) the lower left region are identified as star forming (non-star-forming) gas and are indicated with the blue (red) colour. Using the \textit{KMeans} machine-learning clustering algorithm, the non-star-forming gas particles are further divided into two groups, marked by the dashed (hotter) and dotted (colder) contours, respectively. \textit{Top panels:} number densities of all (red solid line), hotter (black dashed line) and colder (black dotted line) non-star-forming gas of the two haloes as functions of temperature. The vertical red dashed, black dashed and black dotted lines represent the {\sc fof} halo virial temperature and the median temperatures of the two non-star-forming gas groups, respectively.}
\end{figure}

On the other hand, hot gas in the SAM represents a reservoir where the temperature is as high as the halo virial temperature as a result of shock--heating during gas accretion (see Section \ref{sec:SAM_cooling}). Gas within the hot gas reservoir cannot form stars in the SAM. Therefore, in order to define a hot gas component for a galaxy in hydrodynamic simulations, which can be compared with the SAM, we first exclude all potential star forming particles from the galaxy. We plot the temperature distribution of non-star-forming gas particles of the two haloes in the \textit{NOSN\_NOZCOOL\_NoRe} {\smaug} simulation in the top panels of Fig. \ref{fig:temp_nosn_nozcool}. In the more massive halo, three populations\footnote{Supernova feedback and reionization are not included in \textit{NOSN\_NOZCOOL\_NoRe}, which heat the ISM and alter the gas temperature-phase diagram as demonstrated in the companion paper.} can be observed, corresponding to, from right to left, (1) infall hot gas; (2) infall gas that has been through cooling, and the cooling rate decreases significantly when temperature\footnote{In collisional ionization equilibrium, hydrogen is ionized at this temperature (${\sim} 5\times10^4\mathrm{K}$), which, however, is low for helium ionization and appears as a trough on the cooling function curve of primordial gas.} drops to a few $10^4\mathrm{K}$ \citep{Wiersma2009a}; and (3) cold gas which is about to reach the multiphase ISM disc and become dense enough to form stars. However, we can only identify the group of less dense (non-star-forming) cold gas in the less massive halo.

Based on the density--temperature phase diagram of the non-star-forming gas particles of a galaxy, the hot gas mass and temperature of the galaxy can be determined. In practice, depending on the number of non-star-forming gas particles ($N_\mathrm{nonSF}$) that a galaxy comprises,
\begin{enumerate}
	\item when $N_\mathrm{nonSF}=0$, the hot gas mass of the galaxy is assigned 0;
	\item when $N_\mathrm{nonSF}=1$, the hot gas mass and temperature of the galaxy are assigned the properties of the gas particle ($m_\mathrm{nonSF}$ and $T_\mathrm{nonSF}$) if $T_\mathrm{nonSF}>T_\mathrm{crit}$, otherwise, the hot gas mass of the galaxy is assigned 0;
	\item when $N_\mathrm{nonSF}>1$, we use the \textit{scikit-learn} \textsc{python} package \citep{scikit-learn} -- \textit{KMeans}, a clustering algorithm commonly used for machine learning. We consider two features (density and temperature) and separate the non-star-forming gas particles into two groups (see the dashed and dotted contours in the bottom panels of Fig. \ref{fig:temp_nosn_nozcool}).
	\begin{enumerate}
		\item If the median temperatures of the two groups differ by more than $\Delta_\mathrm{T, crit}$ in logarithm, the hot gas mass and temperature of the galaxy are assigned the median temperature and the total mass of the hotter group (the dashed contour in the left bottom panel of Fig. \ref{fig:temp_nosn_nozcool}).
		\item If the median temperatures of the two groups differ less than $\Delta_\mathrm{T, crit}$ in logarithm, the two groups of the non-star-forming gas particles are considered as one cluster. In this case, the hot gas mass and temperature of the galaxy are assigned the total mass and the median temperature ($\bar{T}_\mathrm{nonSF}$) of all non-star-forming gas particles if $\bar{T}_\mathrm{nonSF}>T_\mathrm{crit}$, otherwise, the hot gas mass of the galaxy is assigned 0.
		
		We adopt $T_\mathrm{crit}=10^5\mathrm{K}$ and $\Delta_\mathrm{T, crit}=0.5$ in this work and we note that, for haloes with efficient cooling\footnote{Molecular cooling is not implemented in this work. Therefore only haloes that are more massive than the atomic cooling limit possess efficient cooling.}, the determination of hot gas is not sensitive to the temperature thresholds (i.e. $T_\mathrm{crit}$ and $\Delta_\mathrm{T, crit}$; see Section \ref{sec:gas_infall1} and Appendix \ref{app:sec:hot}).
	\end{enumerate} 
\end{enumerate}
When the properties of hot gas are defined, we also characterize the remaining particles and identify them as the cold non-star-forming gas component, whose temperature and mass are represented by the median temperature and total mass of the remaining particles.

In the top panels of Fig. \ref{fig:temp_nosn_nozcool}, we show the temperature function of the two groups separately and indicate their median temperatures with black lines with line styles corresponding to the two contours in the bottom panels of Fig. \ref{fig:temp_nosn_nozcool}. We also indicate the virial temperature\footnote{The virial temperature is calculated using the {\sc fof} group properties (see Section \ref{sec:SAM_accretion}).} of the system using vertical red dashed lines. We see that the median temperature of the hotter group is in agreement with the virial temperature for the more massive halo, suggesting the infall gas particles of massive haloes have been through virial shocks and are heated to the virial temperature. However, less massive galaxies do not comprise any hot gas component. We will further discuss it in Section \ref{sec:gas_infall1}.

\section{Modifications of halo masses and baryon fractions}\label{sec:halo mass}

Before performing any comparisons, there is one modification regarding the suppressed growth of dark matter haloes that needs to be addressed. Hydrostatic pressure keeps gas from collapsing into shallow potential wells of low-mass haloes, which consequently decreases the gravitational potential within these haloes and slows their growths compared to a collisionless universe. This has been illustrated using comparisons between hydrodynamic simulations and \textit{N}-body simulations of dark matter only \citep{Sawala2013,Schaller2014,Velliscig2014}, which show that the inclusion of baryons significantly reduces halo masses\footnote{{\color{black}We refer to the halo mass as the total mass of dark matter and baryons.}} (e.g. ${\sim}20$ per cent for haloes around $10^{11} \mathrm{M}_\odot$ at $z\sim0$). 

\begin{figure*}
	\centering
	\includegraphics[width=1.\textwidth]{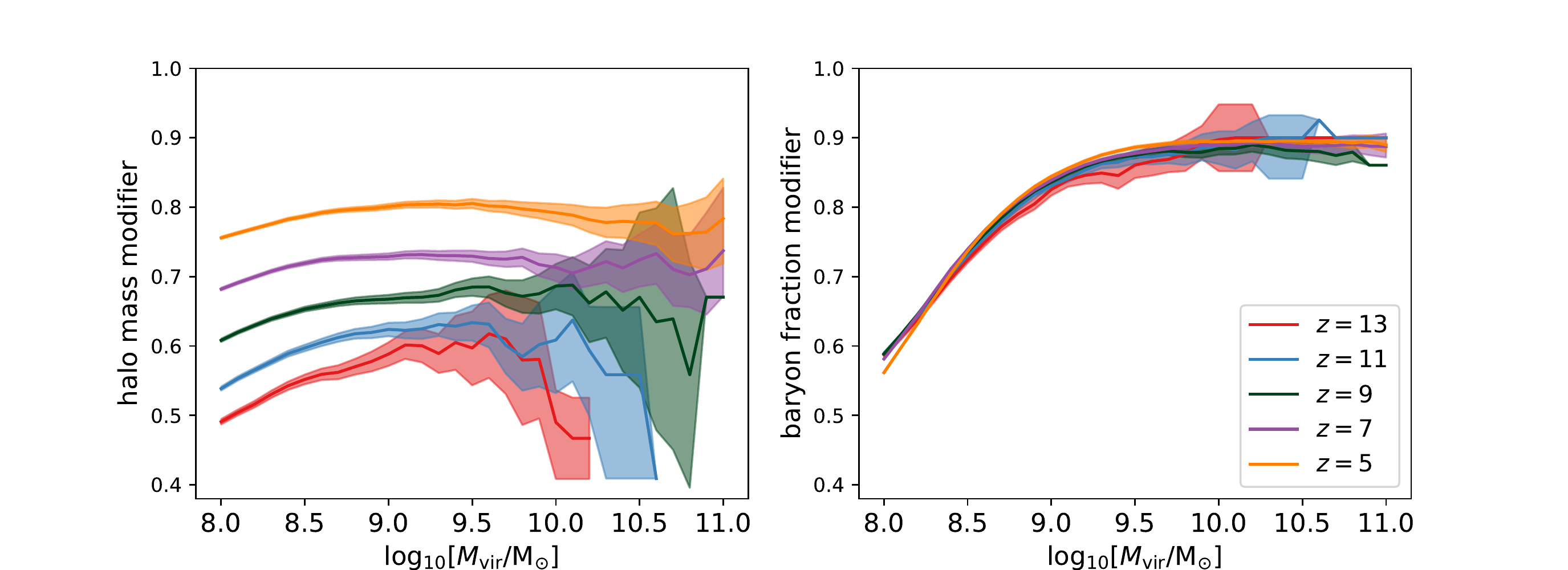}\vspace*{5mm}\\
	\includegraphics[width=\textwidth]{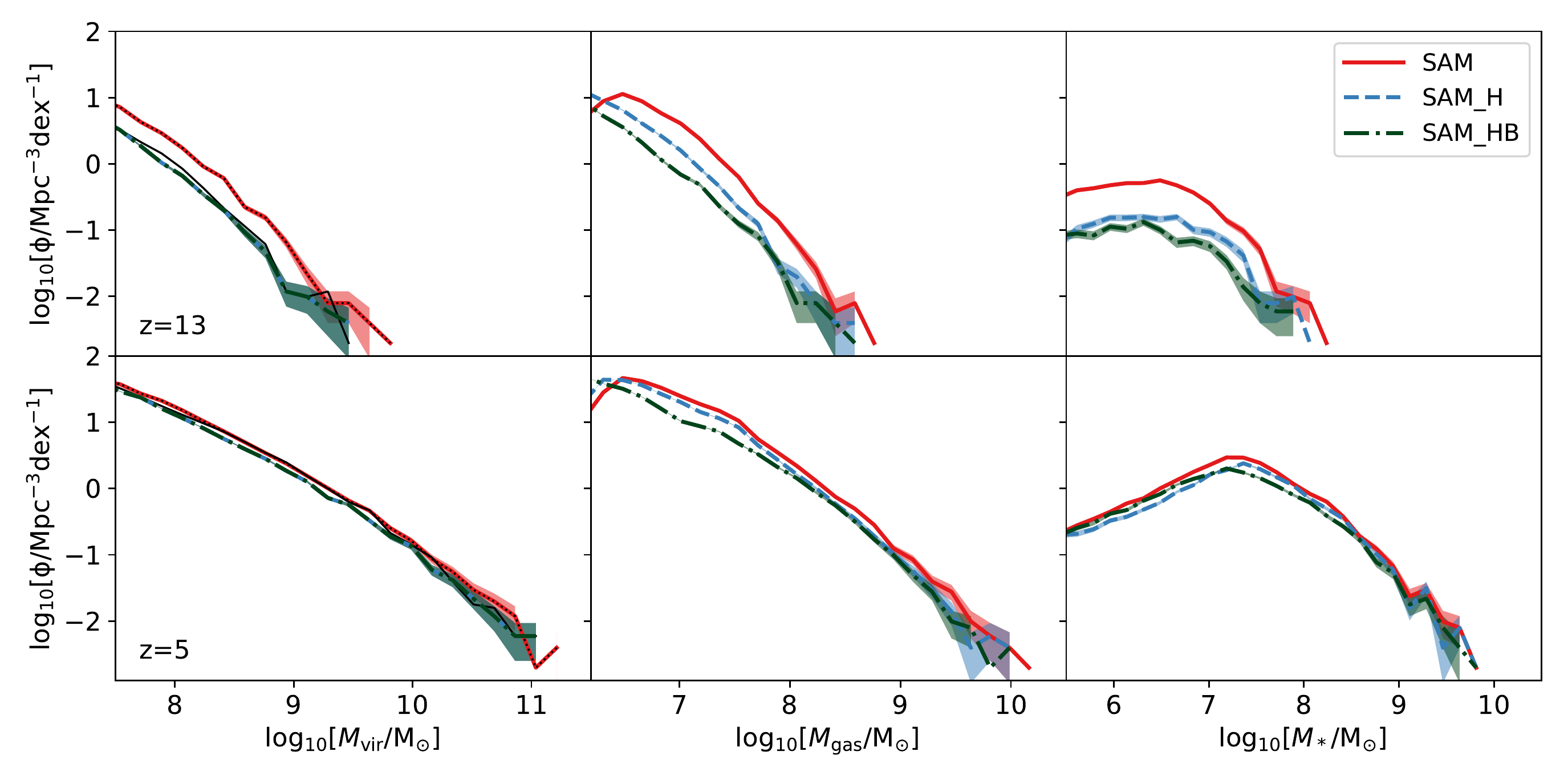}\vspace*{2mm}
	\caption{\label{fig:modifiers}\textit{Top panels:} halo mass modifier and baryon fraction modifier as functions of halo mass in \textit{N}-body simulation at $z=13-5$, with solid lines and shaded regions representing the mean values and uncertainties showing 95 per cent confidence intervals around the mean using 100000 bootstrap resamples. Note that only the means are implemented as modifiers in this work. \textit{Bottom panels:} comparisons of the virial mass, gas mass and stellar mass functions (from left to right) at $z=5$ and 13 between three SAM \textit{NOSN\_NOZCOOL\_NoRe} runs with fixed parameters but different implementations of modifiers: 1) no modifiers (\textit{SAM}); 2) halo mass modifier only (\textit{SAM\_H}); and 3) two modifiers including halo mass and baryon fraction (\textit{SAM\_HB}). Shaded regions represent the $1\sigma$ Poisson uncertainties. For comparison, the halo mass functions predicted by \textit{NOSN\_NOZCOOL\_NoRe} {\smaug} hydrodynamic simulation and the \textit{N}-body simulation are indicated by the thin black solid and dotted lines, respectively.}
\end{figure*}

We investigated the same baryonic effect in dwarf galaxies at high redshift ($z>5$) in \citet[hereafter \citetalias{Qin2017a}]{Qin2017a}. We found that the reduction of mass can be up to a factor of 2 and that the fraction of baryons in haloes with masses between $10^7$ and $10^9 \mathrm{M}_\odot$, which host dwarf galaxies, never exceeds 90 per cent of the cosmic mean ($\Omega_\mathrm{b}/\Omega_\mathrm{m}$) during reionization. Thus, applying SAMs to halo merger trees generated from a collisionless \textit{N}-body simulation and assuming the baryon fraction of every virialized system is $\Omega_\mathrm{b}/\Omega_\mathrm{m}$ overestimates halo and baryon masses of dwarf galaxies at high redshift, and suggests the necessity of incorporating modifications to SAMs. We proposed two modifiers using a simplified hydrodynamic simulation in \citetalias{Qin2017a} -- \textit{ADIAB}, where gas only cools adiabatically with no stellar or galactic physics involved. {\color{black}Note that the halo mass modifier was calculated as the mass ratio of matched haloes (see matching between galaxies in Appendix \ref{sec:match}) between \textit{ADIAB} and \textit{DMONLY} while the baryon fraction modifier is the baryon fraction of these haloes in units of $\Omega_{\mathrm{b}}/\Omega_{\mathrm{m}}$ calculated from the \textit{ADIAB} simulation.} In this work, which aims to inform a better semi-analytic description of high-redshift dwarf galaxies, we make use of these two modifiers, which are shown in the top panels of Fig. \ref{fig:modifiers} as functions of halo mass at $z=13-5$. In practice, the halo mass modifier is included when halo properties are read from the merger trees and the baryon fraction modifier is embedded in $f_\mathrm{mod}$ (see equation \ref{eq:accretion}).

However, we also point out the fact that these modifiers do not offer a consistent modification\footnote{Modifications on subgroup properties are not considered.} accounting for all aspects of astrophysical processes. In \citetalias{Qin2017a}, these have been shown to provide additional ${\sim}10$ per cent alterations to halo mass due to radiative cooling, star formation and supernova feedback. Reionization, on the other hand, provides a more dramatic suppression, but can be incorporated through more efficient reionization feedback, which will be further discussed in the companion paper.

In order to demonstrate the effect of incorporating the modifiers within semi-analytic galaxy formation modelling, we apply {\meraxes} with different implementations. We first apply the SAM with both the halo mass and baryon fraction modifiers implemented and calibrate the free parameters in the \textit{NOSN\_NOZCOOL\_NoRe} regime where feedback is not included (\textit{SAM\_HB}). The detailed calibration strategy is addressed in the next section. Next with all free parameters remaining the same, we then apply the SAM with only the halo mass modifier included (\textit{SAM\_H}) and without any modifiers (\textit{SAM}). We show the halo mass, gas mass and stellar mass functions at $z=5$ and 13 of the three SAM results in the bottom panels of Fig. \ref{fig:modifiers}. The halo mass functions predicted by the {\smaug} simulations are also included in the left subpanels. Note that matching of individual systems (see Appendix \ref{sec:match between meraxes and smaug}) is not performed in this section. We see that compared to the hydrodynamic simulations (black solid lines), the collisionless \textit{N}-body simulation (black dotted lines) as well as the uncorrected SAM overestimate the virial mass function, especially at higher redshifts. However, the halo mass modifier results in a SAM virial mass function that is in better agreement with the hydrodynamic calculation. This leads to a reduction as much as a factor of 2 in of the halo number density at $z\sim12$ for a given mass, and since the growth of the system becomes suppressed, it harbours fewer baryons (see Section \ref{sec:SAM_accretion}) and results in suppressed mass functions of gas and stars. In addition, the baryon fraction modifier due to hydrostatic pressure causes a further reduction of baryons for a given halo mass, which subsequently causes less massive stellar masses.

To summarize, the mass ratio modifier can be interpreted as providing slower evolution of haloes when considering the hydrostatic pressure from baryons, while the baryon fraction modification accounts for the fact that the cosmic mean, $\Omega_\mathrm{b}/\Omega_\mathrm{m}$, cannot be achieved for haloes hosting dwarf galaxies -- gas accretion is less efficient.% In the companion paper, we will show that when supernova feedback is included, the modifications have a less significant impact to the stellar mass function at the range where observational constraints are available.

\section{Comparison between dwarf galaxies in SAM and hydrodynamic simulations }\label{sec:results}
In this section, we present the comparison of accretion and cooling between {\meraxes} and {\smaug}. We note that in this section, if not otherwise stated,

\begin{enumerate}
	\item[(1)] for the purpose of performing fair comparisons of the gas reservoir, we define star-formation gas, non star-formation gas and hot gas in hydrodynamic simulations following the method introduced in Sections \ref{sec:cold gas} and \ref{sec:hot gas};
		
	\item[(2)] for the purpose of facilitating direct comparisons, we match each individual galaxy in the {\meraxes} and {\smaug} outputs using the method presented in Appendix \ref{sec:match between meraxes and smaug}. The free parameters in the SAM, which are summarized in Table \ref{tab:SAMs}, are calibrated to reproduce the $z=5$ stellar mass function predicted by the \textit{NOSN\_NOZCOOL\_NoRe} hydrodynamic simulation;
	
	\item[(3)] for the purpose of minimizing the impact from mismatch and central-satellites switching (see Appendix \ref{sec:gbptrees}), in the SAM, central galaxies with $M_\mathrm{vir}<10^{7.5}\mathrm{M}_\odot$ as well as all satellites\footnote{Satellite haloes in the SAM do not receive any fresh gas, due to the assumption that their central haloes dominate the gravitational potential, which might not be the case in hydrodynamic simulations.} are excluded from the final matched galaxy sample;
	
	\item[(4)] for the purpose of providing more consistent halo properties between {\meraxes} and {\smaug}, the two modifiers (i.e. halo mass and baryon fraction; see \textit{SAM\_HB} in Section \ref{sec:halo mass}) have been implemented in the SAM.
\end{enumerate}

\begin{table*}
	\caption{A summary of different SAM results with values of the main parameters used and modified in this work.}\label{tab:SAMs}
	\begin{threeparttable}
		\begin{tabular}{lccll}
			\hline
			\hline\rule{0pt}{5ex}\vspace*{0.4mm}
			\parbox{20mm}{Name}&
			\parbox{15mm}{Halo mass modifier (Section \ref{sec:halo mass})}& 
			\parbox{23mm}{Baryon fraction modifier (Section \ref{sec:halo mass}, equation \ref{eq:accretion})}& 
			\parbox{26mm}{Star formation efficiency -- $\alpha_\mathrm{sf}$ (Section \ref{sec:sfefficiency}, equation \ref{eq:sfr})}&
			\parbox{29mm}{Maximum cooling factor -- $\kappa_\mathrm{cool}$ (Section \ref{sec:rapid_cooling}, equation \ref{eq:mcool2})}\\			
			\hline% \rule{0pt}{3ex}
			\textit{SAM}&\texttimes&\texttimes&0.05&1\\
			\textit{SAM\_H}&\checkmark&\texttimes&0.05&1\\
			\textit{SAM\_HB}&\checkmark&\checkmark&0.05&1\\
			\textit{SAM\_HBS}&\checkmark&\checkmark&$0.05\times\left(\dfrac{1+z}{6}\right)^{-1.3}$&1\\
			\textit{SAM\_HBSC}&\checkmark&\checkmark&$0.05\times\left(\dfrac{1+z}{6}\right)^{-1.3}$&$\min\left(5, \dfrac{1+z}{6}\right)$\\
			\hline
		\end{tabular}
	\end{threeparttable}
\end{table*}

\begin{figure*}
	\includegraphics[width=\textwidth]{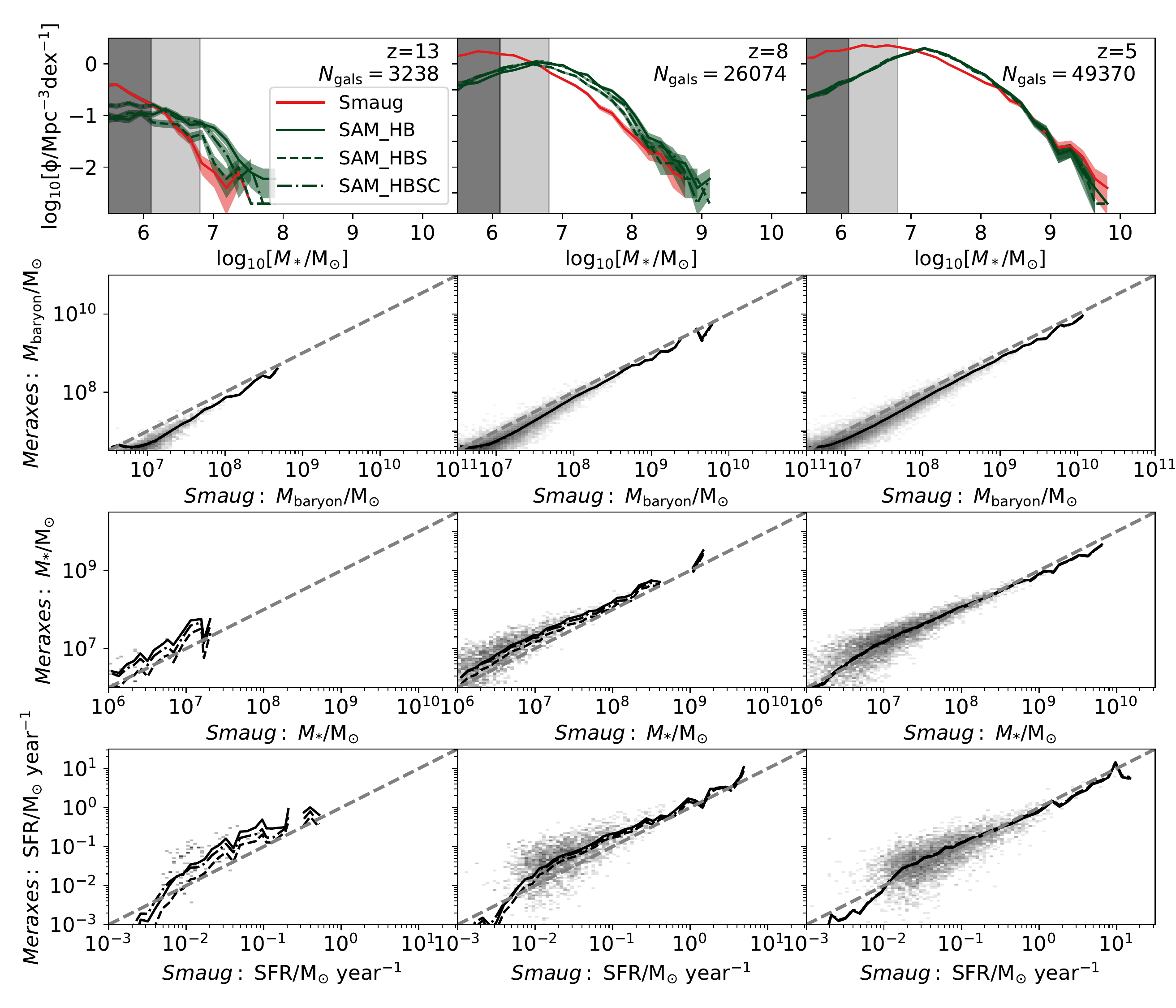}
	\caption{\label{fig:nosn_nozcool_nore}\textit{First row:} the stellar mass functions at $z\sim13-5$ in the \textit{SAM\_HB} and {\smaug} \textit{NOSN\_NOZCOOL\_NoRe} results. Shaded regions represent the $1\sigma$ Poisson uncertainties. The number of matched galaxies, $N_\mathrm{gals}$ is indicated in the top right corner of each panel in the first row. The {\meraxes} results with a redshift-dependent star formation efficiency (\textit{SAM\_HBS}) and redshift-dependent star formation efficiency and cooling rate (\textit{SAM\_HBSC}) are shown with green dashed and dash-dotted lines, respectively. The two grey regions mark the approximate resolution of the simulation, which correspond to 10 and 50 stellar particles, respectively. \textit{Second row:} comparison of the baryon mass between \textit{SAM\_HB} and {\smaug}. The grey 2D histograms indicate the distribution of the entire sample and the black solid lines represent the mean. The grey dashed lines indicate the quantities are identical in the two results. \textit{Bottom two rows:} comparisons of the stellar mass and SFR between {\smaug} and \textit{SAM\_HB}. The comparisons between \textit{SAM\_HBS}, \textit{SAM\_HBSC} and {\smaug} are indicated with dashed and dash-dotted lines, respectively.}
\end{figure*}

The first row of Fig. \ref{fig:nosn_nozcool_nore} presents the stellar mass function of matched galaxies between \textit{SAM\_HB} and {\smaug} at $z=13-5$. We see that the two $z=5$ mass functions are in agreement at $M_*>10^7\mathrm{M}_\odot$, below which the models are limited by resolution and cooling mechanism. This indicates that the two modelled galaxy catalogues comprise similar stellar components at $z=5$ in a cosmological context. However, the results diverge towards higher redshifts -- \textit{SAM\_HB} produces more massive galaxies than {\smaug}. This highlights the issue that stellar build-up proceeds faster in the SAM. As a consequence of the high-redshift modelling, the SAM produces more ionizing photons\footnote{Assuming other relevant factors including the escape fraction remain the same.} at earlier times compared to the hydrodynamic calculation, overestimating the contribution of high-redshift dwarf galaxies to reionization \citep{Liu2016}.

We investigate the properties of individual galaxies in the two models, and show the comparisons of the total baryon mass, stellar mass and SFR predicted by the hydrodynamic simulation and SAM in the bottom three rows of Fig. \ref{fig:nosn_nozcool_nore}. The grey 2D histograms and black solid lines represent the distribution and mean of all matched galaxies, respectively. We see that the overall baryon mass is underestimated by the SAM, with the offset becoming larger at the low-mass end. This is because the \textit{NOSN\_NOZCOOL\_NoRe} {\smaug} simulation predicts larger halo masses and baryon fractions than \textit{ADIAB} (where only adiabatic cooling of gas is included; see Section \ref{sec:halo mass}) due to cooling and star formation (\citetalias{Qin2017a}). Therefore, modifying the halo mass and baryon fraction using the \textit{ADIAB} {\smaug} simulation overestimates the baryonic effect when feedback is not included. In addition, we see that both stellar mass and SFR are similar between the two models at $z=5$, which are, however, underestimated in the SAM result when the stellar mass approaches $10^6\mathrm{M}_\odot$ or SFR reaches $10^{-2}\mathrm{M}_\odot\mathrm{yr}^{-1}$. These objects are close to or below the atomic cooling threshold, where cooling becomes inefficient. For these objects, while a fraction of gas particles in the hydrodynamic simulation are still dense enough to form stars, star formation is quenched in the SAM due to the lack of replenishment of cold gas (see Section \ref{sec:SAM_SF}). This causes an underestimation of the number of ionizing photons at the high redshifts when the first galaxies start to ionize the IGM. 

We next discuss the involved prescriptions that might induce such discrepancies in the history of star formation. We note that in the \textit{NOSN\_NOZCOOL\_NoRe} case when feedback is not implemented, there are only two processes that might cause the different stellar growth histories between SAMs and hydrodynamic simulations: 1) star formation efficiency; and 2) gas fuelling and cooling efficiencies\footnote{{\color{black}In this work, we keep the original cooling functions of primordial elements adopted in \citetalias{Mutch2016a} and \citealt{duffy2010impact}. The difference in the two cooling curves (i.e. {\sc mapping ii} for the SAM, \citealt{Sutherland1993} and {\sc cloudy} for the hydrodynamic simulations, \citealt{ferland1998cloudy}) is minor and does not have a significant impact on our results.}}.

\subsection{Star formation efficiency}\label{sec:sfefficiency}
We note again that since star formation is not resolved in cosmological simulations, both the hydrodynamic simulation and SAM start from an empirical relation between SFR and density -- the KS law \citep{kennicutt1998global}, which proposes that a galaxy is able to form stars when its surface density ($\Sigma_\mathrm{g}$) exceeds a critical value ($\Sigma_\mathrm{crit}{\sim} \mathrm{a\ few\ M_\odot pc^{-2}}$), and the surface SFR density ($\dot{\Sigma}_\star$) can be estimated by
\begin{equation}\label{eq:ks}
\dfrac{\dot{\Sigma}_\star}{\left(2.5\pm0.7\right)\times10^{-4}\mathrm{M_\odot {yr}^{-1} {pc}^{-2}}} =\left(\dfrac{\Sigma_\mathrm{g}}{\mathrm{1M_\odot kpc^{-2}}}\right)^n,
\end{equation}
where $n=1.4\pm0.15$ is suggested by observations \citep{kennicutt1998global}. We have shown that cooling is not well modelled in the SAM when the virial temperature is lower than the atomic cooling threshold, which leads to significant underestimations of the number of low stellar mass and SFR objects in Fig. \ref{fig:nosn_nozcool_nore}. However, this cannot explain the discrepancy of star formation history for more massive objects. In this subsection, we further probe the star formation laws utilized in {\meraxes} and {\smaug}.

\subsubsection{Star formation in hydrodynamic simulations} 
In the hydrodynamic simulation, local 3D densities, $\rho_\mathrm{g}$ can be calculated through smoothing particle mass, $m_\mathrm{g}$ within a certain volume. {\smaug} converts the critical surface density to three dimensions ($\rho_\mathrm{crit}$) assuming a self--gravitating disc. The gas on the disc is determined as multiphase ISM with an effective ratio of specific heats, $\gamma_\mathrm{eff} {=} 4/3$. Then the pressure of the disc can be derived from the equation of state. When the local pressure of a gas particle with a relatively low temperature (${<}10^5\mathrm{K}$) exceeds the critical value (see Section \ref{sec:cold gas}), the particle is considered as a potential star forming region. This infers a SFR of (see more details in \citealt{Schaye2008})
\begin{equation}\label{eq:sfr_smaug_tmp}
\dot{m}_{\star,\mathrm{hydro}} {\equiv} \dfrac{m_\mathrm{g}}{t_\mathrm{g}} {\equiv} m_\mathrm{g} \dfrac{\dot{\Sigma}_\star}{\Sigma_\mathrm{g}}\propto m_\mathrm{g} \Sigma_\mathrm{g}^{\left(n{-}1\right)} \propto m_\mathrm{g} \rho_\mathrm{g}^{0.5\left(n{-}1\right)\gamma_\mathrm{eff}},
\end{equation}
where $t_\mathrm{g} \equiv \Sigma_\mathrm{g}/\dot{\Sigma}_\star$ is the gas depletion time-scale, the third step makes use of equation (\ref{eq:ks}) and the last step assumes that $\Sigma_\mathrm{g}$ is of the order of the Jeans column density in the case of self gravitating disc. With a further assumption that the disc follows an isothermal exponential surface density profile \citep{Schaye2004}
\begin{equation}
\rho_\mathrm{g} = \dfrac{G M_\mathrm{disc}^2}{12\pi c_\mathrm{s}^2R_\mathrm{disc}^4},
\end{equation} 
where $M_\mathrm{disc}\propto M_\mathrm{vir}$, $R_\mathrm{disc}\propto R_\mathrm{vir}$ and $c_\mathrm{s}$ are the mass, radius and sound speed of the disc, respectively, we find that for a given halo mass $M_\mathrm{vir}$, $\rho_\mathrm{g} \propto \left(1+z\right)^4$, and hence
\begin{equation}\label{eq:sfr_smaug}
\dot{m}_{\star,\mathrm{hydro}} \propto \left(1+z\right)^{2\gamma_\mathrm{eff}\left(n-1\right)} \appropto \left(1+z\right)^{1.1}.
\end{equation}

\subsubsection{Star formation in SAMs}\label{sec:sfr_meraxes}
On the other hand, the average SFR of a galaxy in the SAM is calculated using equation (\ref{eq:sfr}). For a given halo mass and with $m_\mathrm{cold}\propto M_\mathrm{vir}$, the redshift dependency of $\dot{m}_{\star,\mathrm{SAM}}$ is approximately
\begin{equation}\label{eq:sfr_meraxes}
\dot{m}_{\star,\mathrm{SAM}} = \alpha_{\mathrm{sf}}\times\dfrac{\mathrm{max}\left(0,m_{\mathrm{cold}}{-}m_{\mathrm{crit}}\right)}{t_{\mathrm{dyn,disc}}} \appropto \left(1+z\right)^{1.5}.
\end{equation}
Note that without supernova feedback in the SAM, the fraction of galaxies that comprise a massive cold gas reservoir (i.e. $m_\mathrm{cold}>m_\mathrm{crit}$) increases towards higher redshifts and in more massive objects. This suggests that the index can be larger than 1.5 on average, especially for less massive haloes.

\begin{figure*}
	\begin{minipage}{\textwidth}
		\centering
		\includegraphics[width=0.495\textwidth]{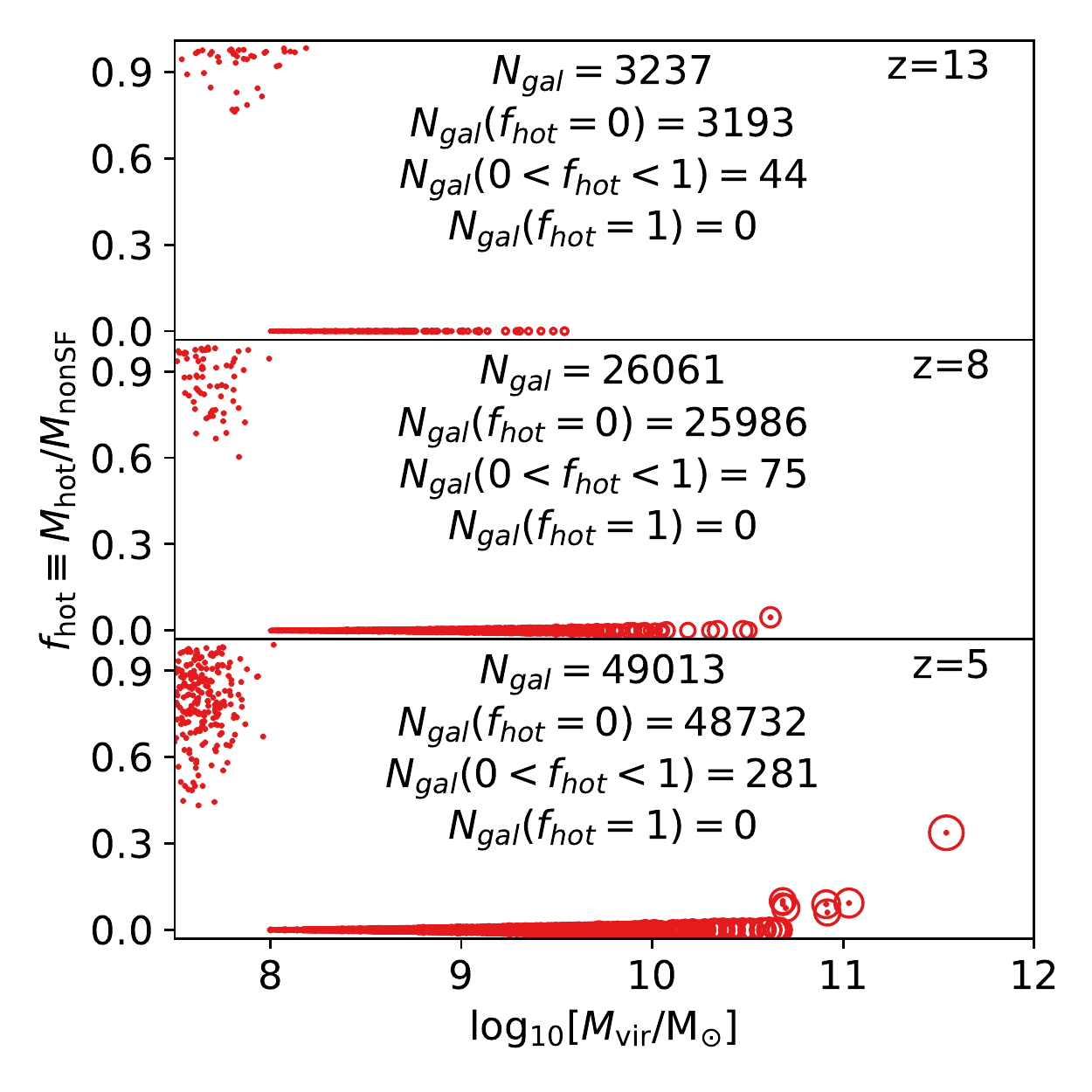}		\includegraphics[width=0.495\textwidth]{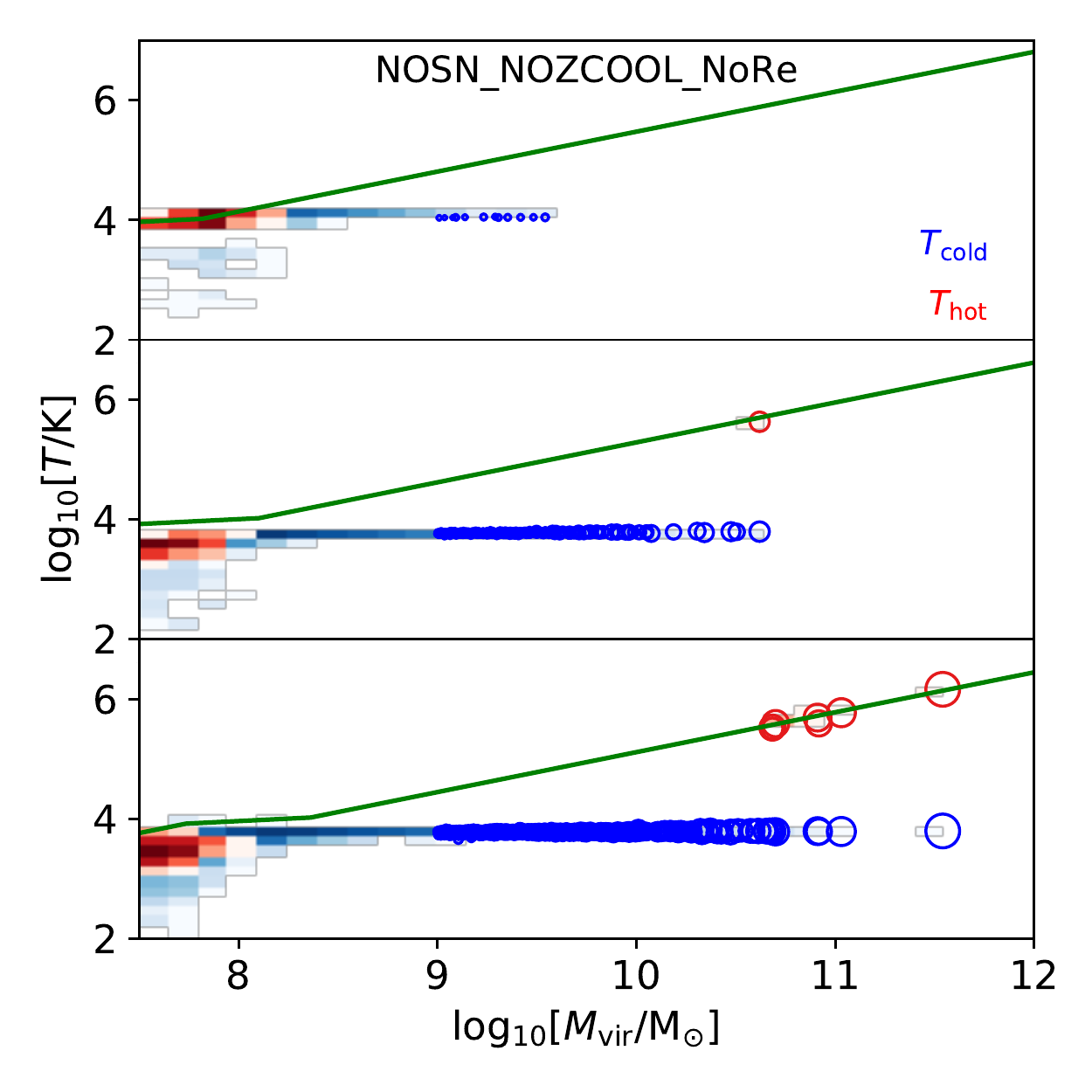}
	\end{minipage}
	%	\vspace*{-3.8mm}
	\caption{\label{fig:temp_nosn_nozcool_nore}\textit{Left panel:} the fraction of hot gas in the non-star-forming gas ($f_\mathrm{hot}\equiv M_\mathrm{hot}/M_\mathrm{nonSF}$) from the \textit{NOSN\_NOZCOOL\_NoRe} {\smaug} simulation at $z=13-5$ as a function of the halo mass ($M_\mathrm{vir}$). Galaxies with $f_\mathrm{hot}\ne0$ are indicated with red dots. Galaxies with $M_\mathrm{vir}>10^8\mathrm{M}_\odot$ are emphasized with red circles with circle size representing the stellar mass ($M_*$). In this mass range, galaxies comprise no or little hot gas. The numbers of all galaxies, galaxies with $f_\mathrm{hot}=0$, $0<f_\mathrm{hot}<1$ and $f_\mathrm{hot}=1$ are shown in each subpanel. \textit{Right panel:} gas temperature as a function of $M_\mathrm{vir}$. {\color{black}The red and blue 2D histograms represent hot and cold non-star-forming gas in the {\smaug} simulation (see Section \ref{sec:hot gas}) while the green solid line indicates the correlation between the virial temperature and halo mass.} Galaxies with $M_\mathrm{vir}>10^9\mathrm{M}_\odot$ are also indicated by the red (hot gas) and blue (cold non-star-forming gas) circles. Fewer than 10 galaxies in our simulation comprise hot gas at $z=5$, and their hot gas has been shock--heated to the virial temperature.}
\end{figure*}

\subsubsection{Redshift-dependent star formation efficiency} 
Comparing equations (\ref{eq:sfr_smaug}) and (\ref{eq:sfr_meraxes}), we find that although both numerical calculations of star formation are derived from the KS law, they possess different evolutionary histories for a given halo mass. This is essentially due to the different assumption of gas depletion time-scale adopted in the two modelling approaches, which is chosen to be the disc dynamical time in the SAM while in the hydrodynamic simulation is inferred from observations. In order to be consistent with the hydrodynamic simulation, a suppressed star formation efficiency of $\alpha_\mathrm{sf} \propto (1{+}z)^{{-}m}$ towards higher redshifts is required in the SAM. {\color{black}We next further discuss the value of $m$.
\begin{enumerate}
	\item The scaling indices in equations (\ref{eq:sfr_smaug}) and (\ref{eq:sfr_meraxes}) assume a disc profile for star-forming gas and a negligible cold gas threshold of star formation (i.e. $m_\mathrm{cold}\gg m_\mathrm{crit}$) in the SAM, respectively. These infer $m\sim0.4$.
	\item However, at higher redshifts, galaxies have larger velocity dispersions, indicating increased turbulence and thickened discs \citep{Newman2012,Price2015}. Meanwhile, simulations suggest that mergers which happen frequently at high redshift \citep{Poole2016} can also thicken discs \citep{Moster2010,Moster2012}. Therefore, the aforementioned assumption of a self-gravitating disc might not be valid at high redshift. If we assume an SIS profile for star-forming gas, $\rho_\mathrm{g}$ scales $\left(1+z\right)^3$, which leads to a larger value of $m\sim0.7$.
	\item Furthermore, during the experiment we find that, in the SAM, galaxies are more likely to have an insufficient cold gas reservoir (i.e. $m_\mathrm{cold}<m_\mathrm{crit}$) at lower redshifts. This requires a more suppressed star formation efficiency at higher redshift, and in practice, $m=1.3$ is adopted in this work.
\end{enumerate}
We note that since the star-forming gas profile varies with different implementations of physics including feedback, $m$ is introduced as a free parameter modulating the global star formation history.}

In Fig. \ref{fig:nosn_nozcool_nore}, the semi-analytic result using the redshift-dependent star formation efficiency (\textit{SAM\_HBS}) is presented using dashed lines. We see that suppressing star formation at higher redshifts results in a better agreement of the stellar mass function, stellar mass and SFR with the hydrodynamic calculations. Note that in the absence of feedback, star formation is not limited by the gas reservoir of galaxies in the SAM (see equation \ref{eq:sfr}). Therefore, updating the star formation efficiency does not have a significant impact on the gas component discussed in the next subsections. 

\subsection{Gas fuelling}\label{sec:gas_infall1}

\subsubsection{Hot gas in hydrodynamic simulations}
We define hot gas fraction as the mass ratio of the hot gas (see Section \ref{sec:hot gas}) to the non-star-forming gas (see Section \ref{sec:cold gas}; $f_\mathrm{hot}\equiv M_\mathrm{hot}/M_\mathrm{nonSF}$), and show its correlation with halo mass in {\smaug} at $z=13-5$ in the left-hand panel of Fig. \ref{fig:temp_nosn_nozcool_nore}. We highlight galaxies above the atomic cooling threshold with red circles (circle size representing the stellar mass) and we see that these galaxies in general comprise little or no hot gas. We show the numbers of all galaxies, galaxies with $f_\mathrm{hot}=0$, $0<f_\mathrm{hot}<1$ and $f_\mathrm{hot}=1$ in each subpanel, and indicate galaxies with $f_\mathrm{hot}\ne0$ with red points. We see that the majority of galaxies at high redshift do not possess any hot gas, and their non-star-forming gas particles are identified as one group -- the cold non-star-forming gas. However, due to the lack of molecular cooling, galaxies with $M_\mathrm{vir}\lesssim10^8\mathrm{M}_\odot$ do comprise two components. In the right-hand panel of Fig. \ref{fig:temp_nosn_nozcool_nore}, we show the 2D histograms of the hot (red) and cold (blue) non-star-forming gas temperatures as functions of halo mass in {\smaug}. We see that for galaxies with $M_\mathrm{vir}\lesssim10^8\mathrm{M}_\odot$, their hot and cold non-star-forming gas particles\footnote{Note that with a smaller threshold of the temperature offset between hot and cold non-star-forming gas ($T_\mathrm{crit}$; see Section \ref{sec:hot gas}), more particles that belong to haloes with $M_\mathrm{vir}\lesssim10^8\mathrm{M}_\odot$ are identified as hot. However, the temperature thresholds (i.e. $T_\mathrm{crit}$ and $\Delta_\mathrm{T,crit}$) have less impact on more massive haloes, and their $f_\mathrm{hot}$ remains zero or small (see Appendix \ref{app:sec:hot}).} possess median temperatures $>10^3\mathrm{K}$ and $<10^3\mathrm{K}$, respectively. This broad distribution on the density--temperature phase diagram is from adiabatic cooling of the gas as the universe expands, which is also indicated by the decreasing median temperature at lower redshifts of these less massive objects.

\subsubsection{Hot gas in SAMs}
Gas in the IGM falls into the gravitational potential of a halo and fuels star formation in the host galaxy (see Section \ref{sec:SAM_accretion}). {\color{black}Although the hydrodynamic simulation and SAM both assume ionization equilibrium for the accreted gas, the temperature profile varies significantly. While gas particles cover a broad range of temperatures from ${\sim}10^3-10^8\mathrm{K}$ \citep{Wiersma2009b} in the hydrodynamic simulation, (hot) gas initially accreted by a galaxy in the SAM is assumed to be shock--heated to the virial temperature (see Section \ref{sec:SAM_cooling}).} Hot gas then cools through thermal radiation and builds a cold gas disc where stars can form (see Section \ref{sec:SAM_SF}). In the \textit{NOSN\_NOZCOOL\_NoRe} regime where metals cooling is not included, temperature alone (assuming the helium fraction is a constant) determines the cooling rate (see equation \ref{eq:t_cool}) and therefore alters star formation in each galaxy. In order to test the validity of this simplified gas infall prescription for modelling of dwarf galaxies, we compare the temperature of non-star-forming gas in the hydrodynamic simulation to the halo virial temperature. In the right-hand panel of Fig. \ref{fig:temp_nosn_nozcool_nore}, we highlight galaxies with $M_\mathrm{vir}>10^9\mathrm{M}_\odot$ in {\smaug} with circles and show the correlation between virial temperature and halo mass with green solid lines. We see that ${\lesssim}10$ galaxies with $M_\mathrm{vir}>10^9\mathrm{M}_\odot$ in the {\smaug} simulation comprise hot gas at $z=5$, and their hot gas temperatures are close to the virial temperatures, suggesting that theses galaxies have been heated through shocks. However, the non-star-forming gas of most galaxies in {\smaug} is considered as cold with $T\lesssim10^4K$, which is much lower than the virial temperature and becomes more common at higher redshifts. {\color{black}This is consistent with the prediction of EAGLE simulations \citep{Schaye2014}, where the mass ratio of hot\footnote{In \citet{Correa2017}, hot gas is defined as particles with cooling time longer than the dynamical time.} to all baryons is less than 1 per cent on average for haloes around $10^{10}\mathrm{M}_\odot$ \citep{Correa2017}, with the fraction decreasing towards higher redshifts and in less massive haloes.}

\subsubsection{Gas accretion in cold mode}

\begin{figure*}
	\begin{minipage}{\textwidth}
		\centering
		\includegraphics[width=0.495\textwidth]{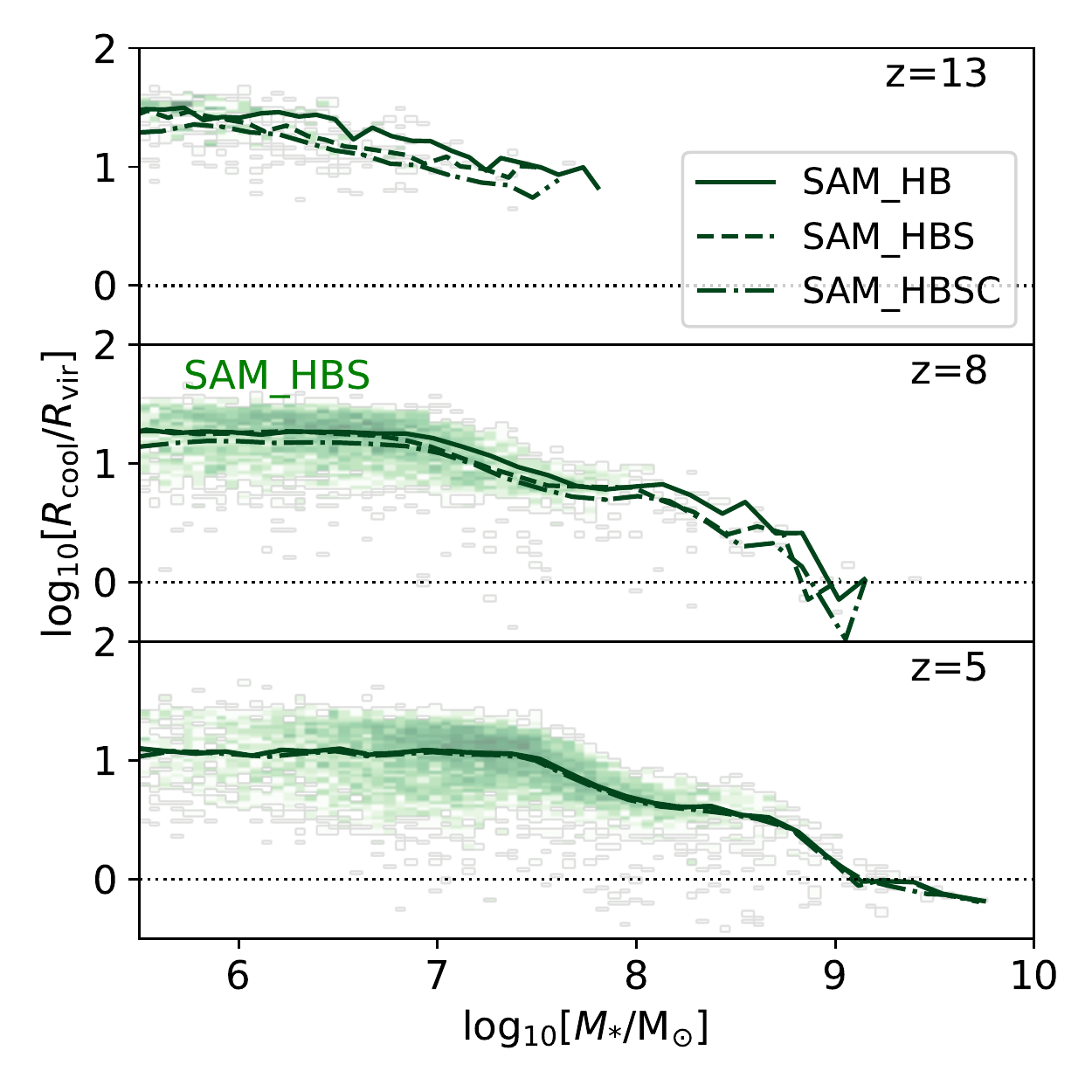}		\includegraphics[width=0.495\textwidth]{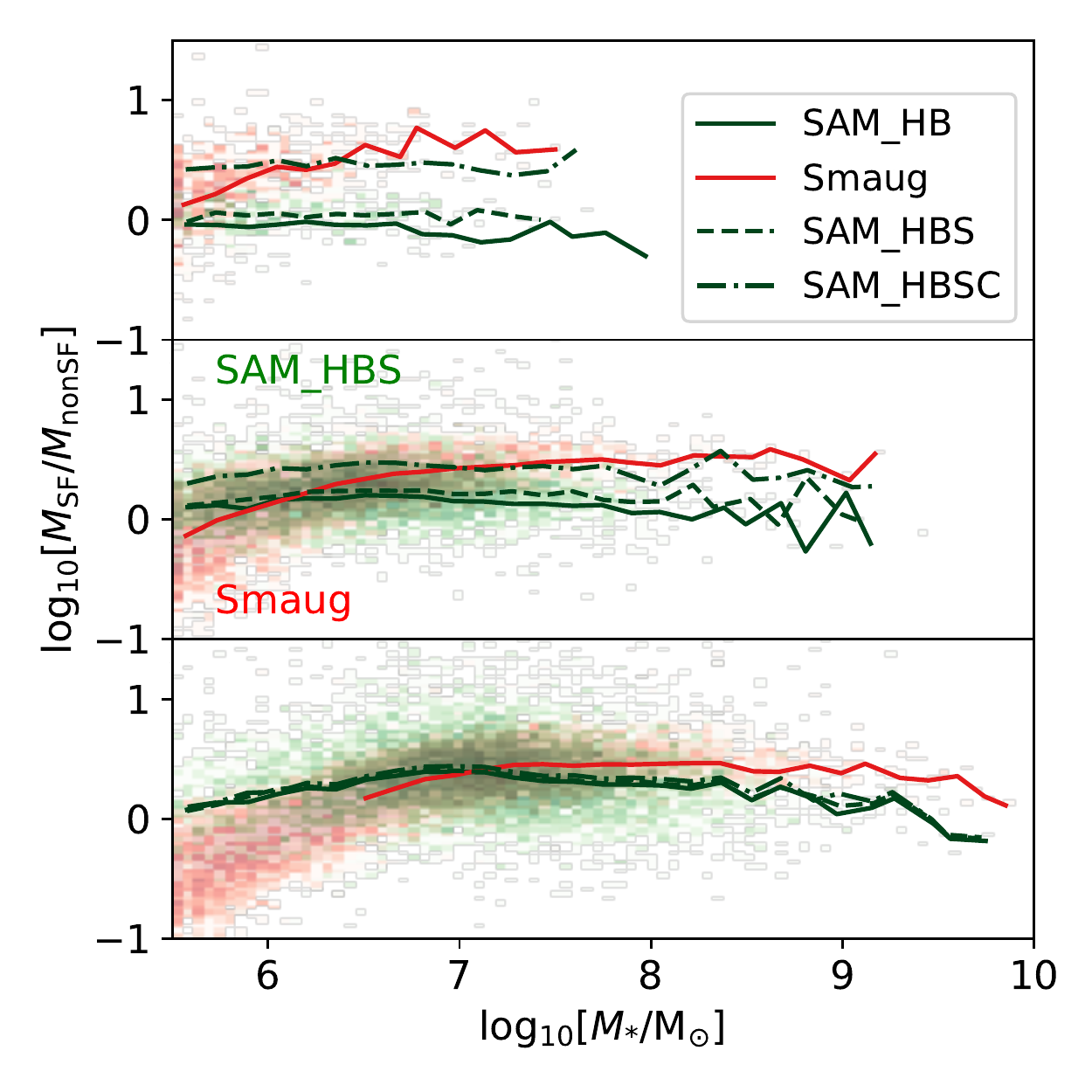}	
	\end{minipage}
	%	\vspace*{-3.8mm}
	\caption{\label{fig:sf_nosn_nozcool_nore}\textit{Left panel:} the ratio of the cooling radius to the virial radius as a function of stellar mass ($M_*$) from the \textit{SAM\_HB} (solid line), \textit{SAM\_HBS} (dashed line) and \textit{SAM\_HBSC} (dash-dotted line) results at $z=13-5$. Lines represent the median while the 2D histogram shows the distribution in \textit{SAM\_HBS}. \textit{Right panel:} the mass ratio of star-formation gas to non-star-forming gas ($M_\mathrm{SF}/M_\mathrm{nonSF}$) as a function of $M_*$. The red and green 2D histograms represent galaxies in the {\smaug} and \textit{SAM\_HBS} results, with the medians shown with red solid and green dashed lines, respectively. The median $M_\mathrm{SF}/M_\mathrm{nonSF}$ from \textit{SAM\_HB} and \textit{SAM\_HBSC} are shown with green solid and dash-dotted lines for comparison. }
\end{figure*}

We note that in the SAM, cold and hot gas, where stars can and cannot form, are defined geometrically. They represent a star forming disc and the outer region, following an exponential and an SIS profile, respectively. However, we show that full virialization of infall gas might be problematic for high-redshift modelling of dwarf galaxy formation. In these less massive systems, virial shocks might not form and the infall gas could then only reach the virial temperature when it arrives at the star forming disc \citep{Birnboim2003,Cattaneo2017}. Instead, these cold and less dense non-star-forming gas particles of dwarf galaxies in {\smaug} represent a cold accretion mode \citep{Keres2005,Keres2009,benson2011,Correa2017}. Therefore, in the SAM, instead of being fully virialized, the infall gas of dwarf galaxies should be separated into hot and cold components, which reach the disc and contribute to star formation on different time-scales (i.e. thermal cooling time, $t_\mathrm{cool}$ in equation \ref{eq:t_cool}, and dynamical time, $t_\mathrm{dyn}$ in equation \ref{eq:mcool}).

In addition, since the non-star-forming gas particles of the majority of high-redshift dwarf galaxies are \textit{not} hot in the hydrodynamic simulation, we advocate that the terminology of hot gas becomes misleading in the redshift and mass ranges discussed in this work.

\subsection{Rapid cooling}\label{sec:rapid_cooling}

The gas infall and cooling prescription adopted for the SAM distinguishes two regimes -- a static hot halo and rapid cooling gas (see Section \ref{sec:SAM_cooling}). In the second regime, gas cools onto the disc at the dynamical time-scale, which has the properties of cold accretion if one combines the process of gas infall and rapid cooling. Therefore, altering the rapid cooling rate is an alternative to incorporating cold accretion\footnote{We leave the task of implementing cold accretion in a future project when molecular cooling is included.} directly \citep{Cattaneo2017}.

The relative importance of the two cooling modes is determined by the ratio of the cooling radius to the virial radius ($R_\mathrm{cool}/R_\mathrm{vir}$), which is shown in the left panel of Fig. \ref{fig:sf_nosn_nozcool_nore}. We see that galaxies with $M_* < 10^9\mathrm{M}_\odot$ are mostly in the rapid cooling regime with $R_\mathrm{cool}/R_\mathrm{vir}>1$. The determination of static hot halo and rapid cooling regimes has been discussed in \citet{croton2006many}. Comparing with hydrodynamic simulations, they found $M_\mathrm{vir}\sim10^{11}\mathrm{M}_\odot$ separates the two regimes, which is approximately independent with redshift up to $z\sim6$ (see also \citealt{Correa2017}). This is in agreement with our finding\footnote{Although we are now focusing on a simplified scenario where supernova and metals are not included, it is also true when the feedback is implemented as we will show in the companion paper.} of $M_* \sim 10^9\mathrm{M}_\odot$.

The balance of instantaneous cooling rate between the two numerical approaches can be inferred from the evolution of the mass ratio of star forming to non-star-forming gas (see Appendix \ref{app:sam_properties} for more information of galaxy properties in the SAM). This is shown in the right panel of Fig. \ref{fig:sf_nosn_nozcool_nore}. We see that while the \textit{SAM\_HBS} result is in agreement with the hydrodynamic calculation at low redshift, $M_\mathrm{SF}/M_\mathrm{nonSF}$ in the SAM is underestimated at higher redshifts. We have shown that in the SAM, dwarf galaxies at high redshift are identified as being in the rapid cooling regime. Therefore, in order to understand the different transition rates between non-star-forming reservoir and star-formation gas in {\meraxes} and {\smaug}, we next discuss the assumed SIS profile of hot gas reservoir in the SAM.

\subsubsection{Gas profile}

\begin{figure*}
	\begin{minipage}{\textwidth}
		\centering
		\includegraphics[width=0.495\textwidth]{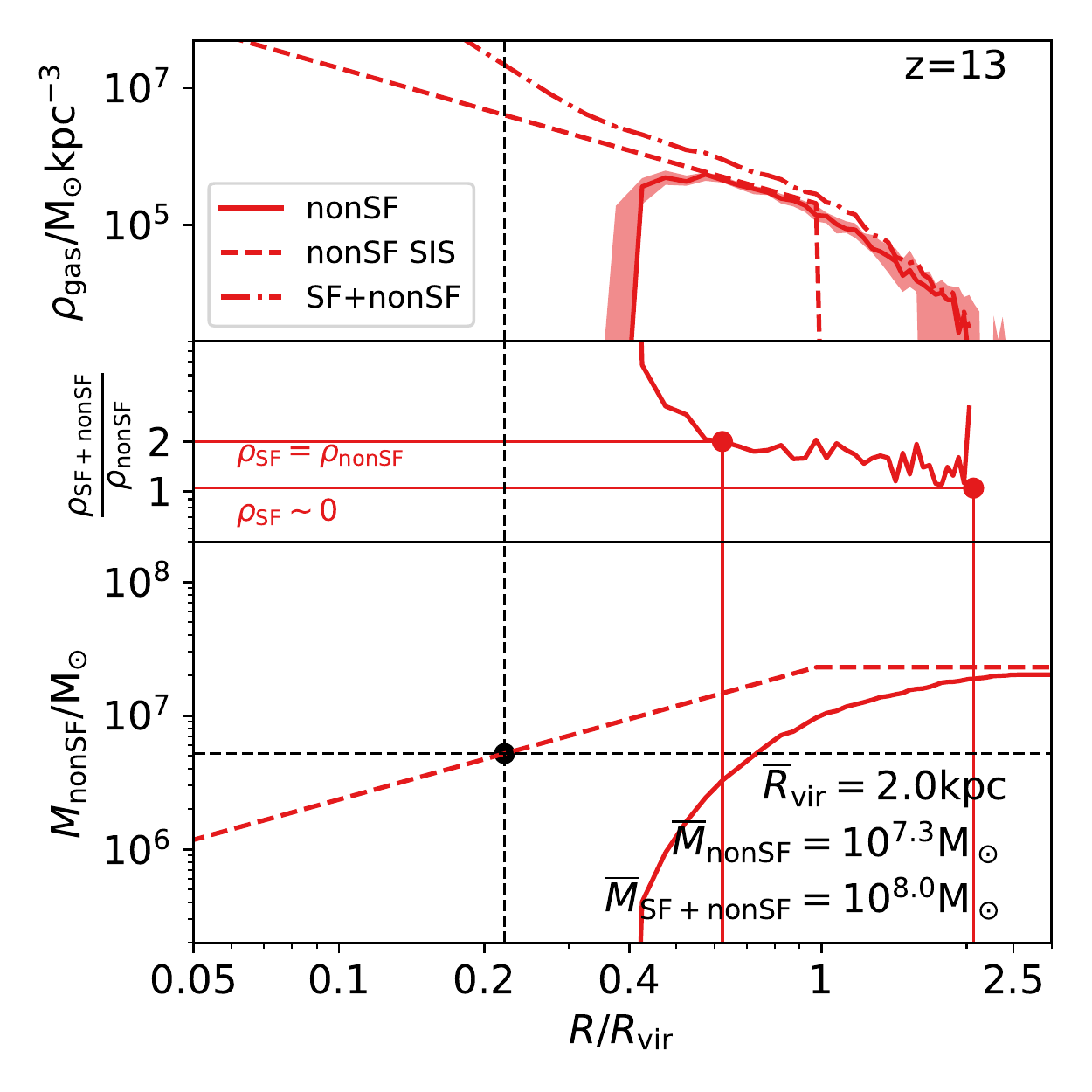}	\includegraphics[width=0.495\textwidth]{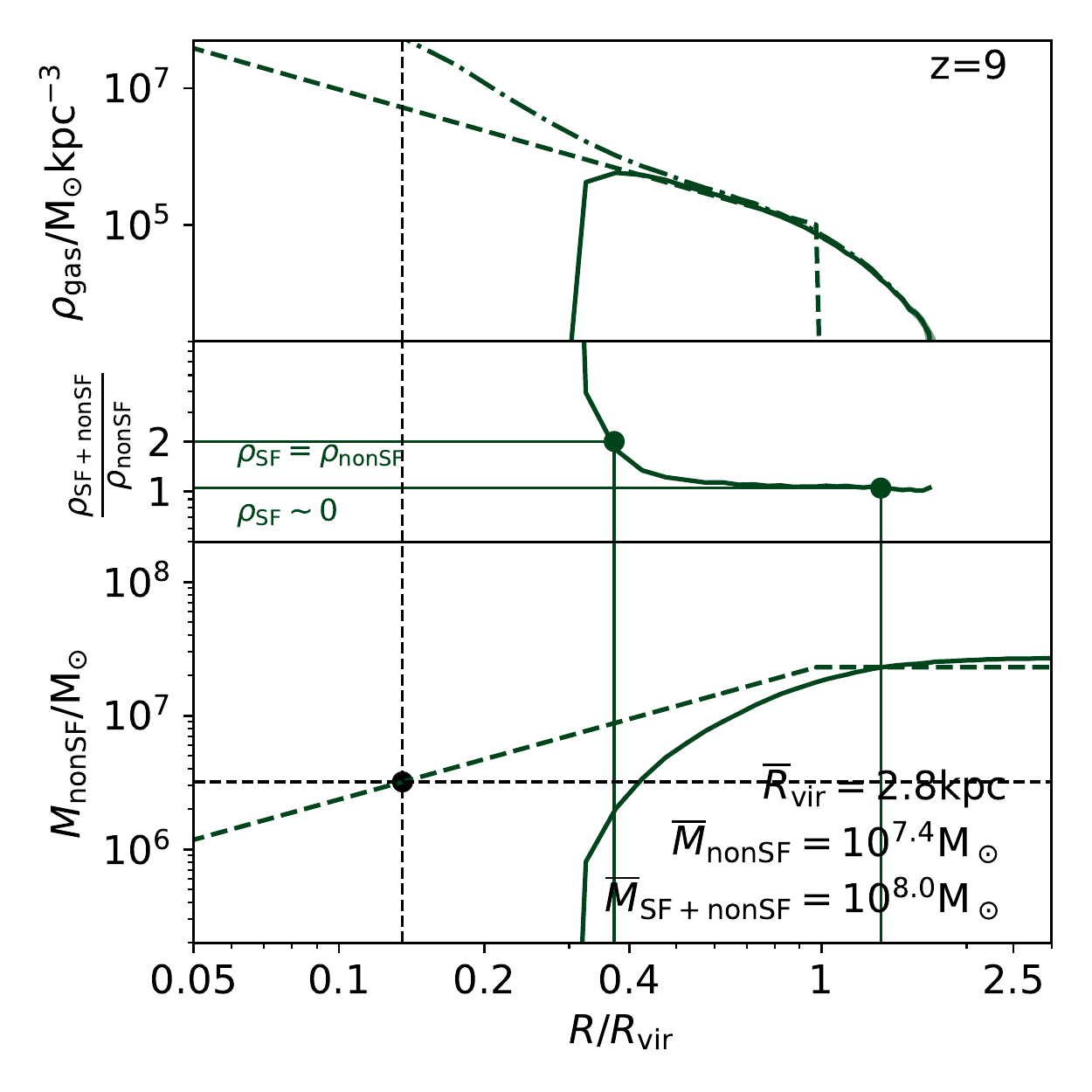}\\%	\vspace*{-3mm}
		\includegraphics[width=0.495\textwidth]{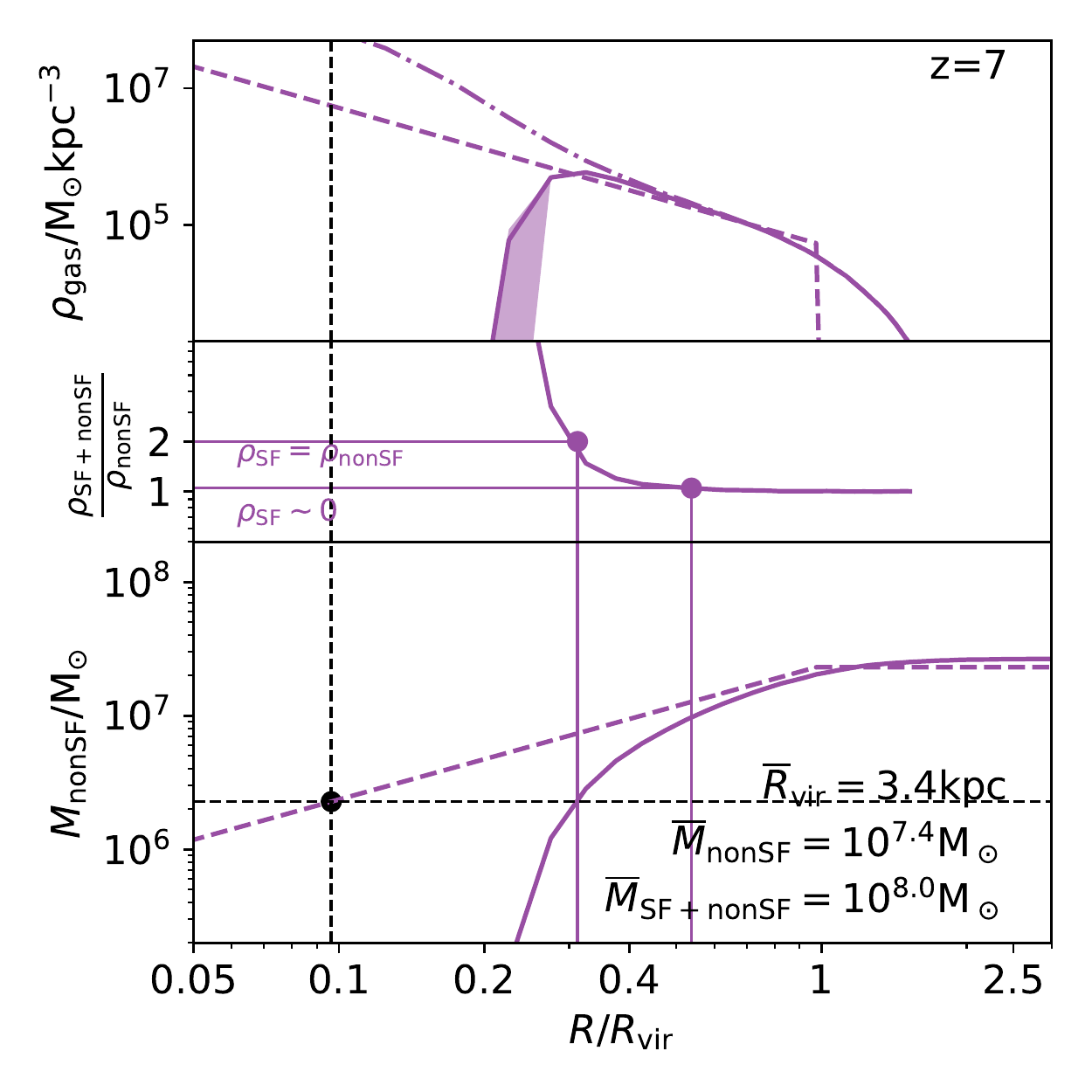}		\includegraphics[width=0.495\textwidth]{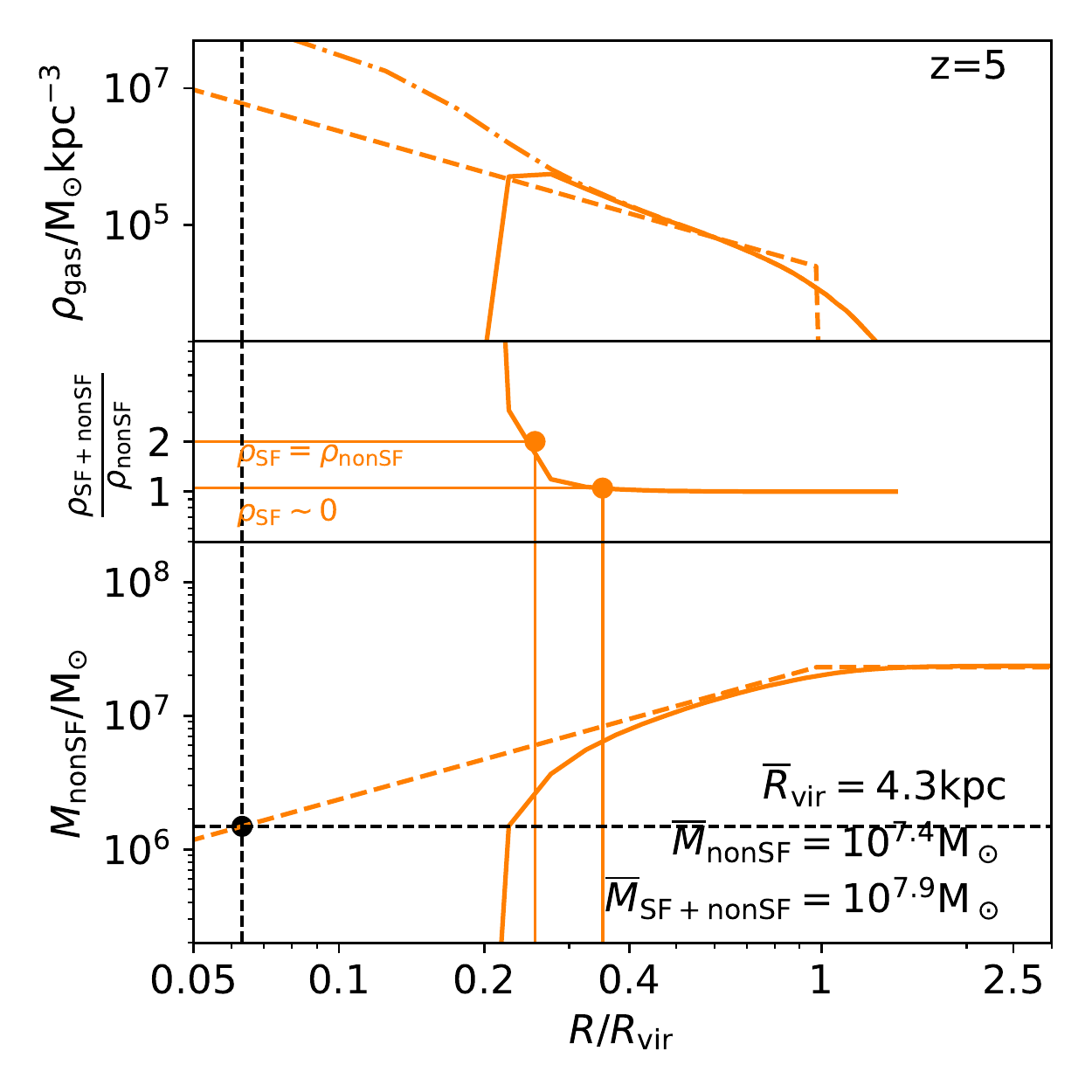}%\vspace*{-3.5mm}
	\end{minipage}
	\caption{\label{fig:1dprofile}Gas profiles of galaxies with $M_*{\sim}10^{7\pm0.5}\rm{M}_\odot$ at $z=13-5$ in the \textit{NOSN\_NOZCOOL\_NoRe} {\smaug} simulation (coloured lines). In each panel, \textit{top:} the median radial density profiles of all gas (i.e. star forming and non-star-forming gas; dash-dotted line) and the non-star-forming gas (solid line). Lines with shaded regions represent the median and 95 per cent confidence intervals around the median using 100 000 bootstrap resamples of the non-star-forming gas profile. The median SIS profile assumed in the SAM is calculated using the same amount of non-star-forming gas, and is indicated with the coloured dashed line; \textit{middle:} the ratio of the density profiles of all gas to non-star-forming gas; \textit{bottom:} the median radial mass profile of non-star-forming gas (solid line) compared to the SIS profile (dashed line). The radius, within which the SIS gas is able to reach the centre through free-fall after one time step in the SAM, and the enclosed SIS gas mass are indicated with thin vertical and horizontal dashed lines, respectively. The radii, where the star-formation gas is as dense as the non-star-forming gas (i.e. $\rho_\mathrm{SF}=\rho_\mathrm{nonSF}$) or becomes deficient (i.e. $\rho_\mathrm{SF}=0$), are indicated with thin solid lines. The median virial radius, masses of non-star-forming, and all gas mass are shown in the bottom right corner of each panel.}
\end{figure*}

We present the median radial profile of the non-star-forming gas of galaxies with $M_*=10^{7\pm0.5}\mathrm{M}_\odot$ in {\smaug} at $z=13-5$ in Fig. \ref{fig:1dprofile}. We see that the density of non-star-forming gas drops at the inner region (where particles become star forming) when compared to an SIS mass profile that follows \citep{croton2006many}
\begin{equation}
\mathscr{M}_\mathrm{nonSF}\left(r\right) = M_\mathrm{nonSF}\times \min\left(\dfrac{r}{R_\mathrm{vir}},1\right),
\end{equation}
the non-star-forming gas particles 1) have a large dispersion up to ${\gtrsim}2R_\mathrm{vir}$; and 2) are less concentrated at higher redshifts. The SAM assumes the SIS profile for hot gas. Therefore, in the rapid cooling regime (see equation \ref{eq:mcool}), the cooling mass during one time step ($\Delta t$) can be calculated through
\begin{equation}
M_\mathrm{cool} = M_\mathrm{nonSF}\times\dfrac{\Delta t}{t_\mathrm{dyn}}\equiv \mathscr{M}_\mathrm{nonSF}\left(r=R_\mathrm{vir}\times\dfrac{\Delta t}{t_\mathrm{dyn}}\right).
\end{equation}
The last step implies there is a radius, within which gas is able to reach the \textit{centre} through free-fall. This is indicated with vertical dashed lines in Fig. \ref{fig:1dprofile} at different redshifts. For comparison, we also present the density profile of all gas of the galaxies with $M_*=10^{7\pm0.5}\mathrm{M}_\odot$ in Fig. \ref{fig:1dprofile} (coloured dash-dotted lines), and indicate the radii, at which the star-formation gas is as dense as the non-star-forming gas (i.e. $\rho_\mathrm{SF}=\rho_\mathrm{nonSF}$) or becomes deficient (i.e. $\rho_\mathrm{SF}\sim0$), using vertical solid lines. We see that the star-formation gas also possesses a large dispersion. At high redshift ($z>9$), $r(\rho_\mathrm{SF}=0)$ can be larger than the virial radius.

\subsubsection{Transition radius of gas reservoir}

We see that the SIS profile significantly overestimates the non-star-forming gas density in the inner region compared to the simulation result. Therefore, with the assumption that gas can only transform from non-star-forming to star forming (or hot to cold in the SAM) gas when collapsing into the centre, the SIS profile leads to an overestimation of the inward collapse rate (see the horizontal dashed line in Fig. \ref{fig:1dprofile}). However, due to the large radial region of star-formation gas observed in Fig. \ref{fig:1dprofile}, this assumption (i.e. transition radius is ${\sim}0$) might not be accurate. Gas particles in dwarf galaxies at $z>9$ can be triggered as star forming regions as far as the virial radius. The combination of corresponding effects between the overestimated collapse rate from the assumed SIS profile and the underestimated transition radius of gas reservoir is the primary cause of the discrepancy in $M_\mathrm{SF}/M_\mathrm{nonSF}$ between {\meraxes} and {\smaug}.

\subsubsection{Redshift-dependent maximum cooling factor}
Whilst adopting the correct gas profile and finding the accurate transition radius of gas reservoir can help explain the underestimation of $M_\mathrm{SF}/M_\mathrm{nonSF}$ in the SAM at high redshift, an analytic solution is not well defined. The large dispersion in the gas profile is mainly due to frequent mergers at these redshifts, which can be affected by numerical configurations, such as particle resolution, as well as by physics implementations including feedback. In order to take the underestimated cooling rate of dwarf galaxies at high redshift into account, we propose an alteration to equation \ref{eq:mcool} as follows
\begin{equation}\label{eq:mcool2}
\dot{m}_{\mathrm{cool}}{=}\dfrac{m_{\mathrm{hot}}}{t_\mathrm{dyn}}\times\min\left(\kappa_\mathrm{cool},\dfrac{r_{\mathrm{cool}}}{R_{\mathrm{vir}}}\right),
\end{equation}
and allow the free parameter, $\kappa_\mathrm{cool}$ representing the maximum cooling factor to exceed unity at high redshift\footnote{The other interpretation of $\kappa_\mathrm{cool}$ is to modulate the time-scale (i.e. $\mathrm{t}_\mathrm{inflow}\equiv\kappa_\mathrm{cool}^{-1}t_\mathrm{dyn}$) of gas inflow from the circumgalactic medium to the ISM. See more in the companion paper.}. We show the result of $M_\mathrm{SF}/M_\mathrm{nonSF}$ calculated with $\kappa_\mathrm{cool}=\min\left(5, 
\dfrac{1+z}{6}\right)$ (\textit{SAM\_HBSC}) in Fig. \ref{fig:sf_nosn_nozcool_nore}, and we see that with higher cooling rates, the semi-analytic calculation becomes more consistent with the hydrodynamic result at high redshift. However, the stellar mass function becomes higher at $z=13$ compared to the {\smaug} result (see Fig. \ref{fig:nosn_nozcool_nore}), suggesting the necessity of stronger suppression of the star formation efficiency (i.e. $m>1.3$) due to the relatively larger disc mass at higher redshifts in the SAM (see Section \ref{sec:sfr_meraxes}).

Lastly, we point out that without a self-consistent radiative transfer calculation, atomic cooling and gas temperature might not be properly simulated in the hydrodynamic simulation either, which alters the importance of thermal radiation cooling. For instance, with local sources (of each gas particle) of ionizing radiation neglected, the cooling rate is potentially overestimated \citep{Schaye2014}, and since self-shielding cannot be properly captured in our hydrodynamic simulations, the cooling rate might be underestimated in dense regions \citep{McQuinn2011}. Moreover, supernova feedback and reionization are expected to regulate galaxy formation through altering the gas component, which will be the topic of a forthcoming paper.

\section{Conclusions}\label{sec:conclusion}
While most assumptions adopted for semi-analytic galaxy formation models, including the gas reservoirs (i.e. the density profile, temperature and transition), are in good agreement with hydrodynamic simulations for Milky Way size objects \citep{Guo2016,Stevens2016b}, they become less accurate for less massive galaxies at high redshift. In this work, we propose modifications to SAMs based on the comparison of dwarf galaxy properties calculated by the {\meraxes} SAM and the {\smaug} high-resolution hydrodynamic simulation. We focus on gas accretion, cooling and star formation at $z\ge5$, and consider scenarios in the absence of reionization and supernova feedback. We summarize the modifications below:

\begin{enumerate}
	\item[(1)] The parent cosmological simulation of a SAM usually includes only collisionless particles, where baryonic physics is neglected. In making comparisons between \textit{N}-body and hydrodynamic simulations starting from identical initial conditions, we previously showed (\citetalias{Qin2017a}) that dwarf galaxy host halo masses are significantly overestimated when hydrostatic pressure is not considered, and that the fraction of baryons accreted by dwarf galaxies cannot reach the level of cosmic mean. While inclusion of halo masses directly from \textit{N}-body dark matter only simulation and assumption of a universal baryon fraction are standard features of SAMs, the impact of these assumptions becomes significant at high redshift and small scale. We have considered implementations to modify the relevant properties (i.e. the halo mass and baryon fraction) in SAMs (see Section \ref{sec:halo mass}).
	
	\item[(2)] We find that the star formation prescription in the SAM that is based on the consumption of a cold gas reservoir does not represent the evolutionary path followed by gas in the hydrodynamic simulation, while both of the two modelling approaches start from the observational relation between surface density and SFR \citep{kennicutt1998global}. We find that this results from variation in the calculated depletion time-scale of the gas reservoir. We address this by modifying the efficiency that modulates star formation in the SAM with a redshift dependency (i.e. replacing the constant $\alpha_{\mathrm{sf}}$ to $\alpha_{\mathrm{sf}}\left(z\right)$)
	\begin{equation}
	\dot{m}_* = \alpha_{\mathrm{sf}}\left(z\right)\times \dfrac{m_\mathrm{cold}-m_\mathrm{crit}}{t_{\mathrm{dyn,disc}}}.
	\end{equation}
	In this work, we adopt $\alpha_{\mathrm{sf}}\left(z\right)=0.05\times\left[\left(1+z\right)/6\right]^{-1.3}$, which allows the model to follow the \textit{NOSN\_NOZCOOL\_NoRe} hydrodynamic simulation in the {\smaug} suite (see Section \ref{sec:sfefficiency}).
	
	\item[(3)] The majority of dwarf galaxies at high redshift are in the rapid cooling regime, where the infalling gas cannot form stable shocks or remain in hydrostatic equilibrium. This represents a cold accretion mode, which is not well modelled with the assumption that the hot gas reservoir follows the SIS profile and falls into the centre within the dynamical time. We find, in the hydrodynamic simulation, that gas in high-redshift dwarf galaxies can form stars as far as the virial radius and that the SIS profile overestimates the density in the inner regions of these low-mass objects. In order to take this into account, we propose of modulation of the cooling prescription with a redshift-dependent collapse rate (i.e. changing the cooling rate upper limit from $m_\mathrm{hot}t_\mathrm{dyn}^{-1}$ to $\kappa_\mathrm{cool}\left(z\right)m_\mathrm{hot}t_\mathrm{dyn}^{-1}$)
	\begin{equation}
	\dot{m}_{\mathrm{cool}}{=}\dfrac{m_{\mathrm{hot}}}{t_\mathrm{dyn}}\times\min\left[\kappa_\mathrm{cool}\left(z\right),\dfrac{r_{\mathrm{cool}}}{R_{\mathrm{vir}}}\right],
	\end{equation}
	where $\kappa_\mathrm{cool}\left(z\right)=\min\left[5, \left(1+z\right)/6\right]$ leads to an agreement with the hydrodynamic calculation on the cold gas mass evolution (see Section \ref{sec:rapid_cooling}).
\end{enumerate}

Furthermore, we point out that the terminology of \textit{hot gas} and \textit{cold gas} becomes misleading when applying SAMs to the formation of high-redshift dwarf galaxies. The hot gas representing non-star-forming gas within a galaxy does not experience full virialization and possess a median temperature that is much lower than the virial temperature of the host halo (Section \ref{sec:gas_infall1}). In a future paper (Qin et al. in prep.), we will discuss star formation in the presence of feedback from reionization and supernovae. We will compare the SAM calculation with the most complete hydrodynamic simulation in the {\smaug} suite that considers reionization as an instantaneous heating background \citep{Haardt2001} and distributes supernova energy in thermal form \citep{DallaVecchia2012}. We will also consider an additional star formation prescription to the one shown in this work, in which molecular gas directly drives the star formation history \citep{Lagos2011}. The goal of these two papers will be more complete and faithful recreation of hydrodynamic simulations at high redshift, which will further leverage the observations that we can make in this early history of our Universe.

\section*{Acknowledgements}
This research was supported by the Victorian Life Sciences Computation Initiative, grant ref. UOM0005, on its Peak Computing Facility hosted at the University of Melbourne, an initiative of the Victorian Government, Australia. Part of this work was performed on the gSTAR national facility at Swinburne University of Technology. gSTAR is funded by Swinburne and the Australian Governments Education Investment Fund. This research was conducted by the Australian Research Council Centre of Excellence for All Sky Astrophysics in 3 Dimensions (ASTRO 3D), through project number CE170100013. This work was supported by the Flagship Allocation Scheme of the NCI National Facility at the ANU, generous allocations of time through the iVEC Partner Share and Australian Supercomputer Time Allocation Committee. AM acknowledges support from the European Research Council under the European Union's Horizon 2020 research and innovation program (Grant No. 638809 -- AIDA). 
\bibliographystyle{\dir mn2e}
\bibliography{reference}

\appendix
\section{Matched Galaxies between \textsc{Meraxes} and Smaug}\label{sec:match}

\begin{figure*}
	\begin{minipage}{\textwidth}
		\subfigure{\includegraphics[width=0.5\columnwidth]{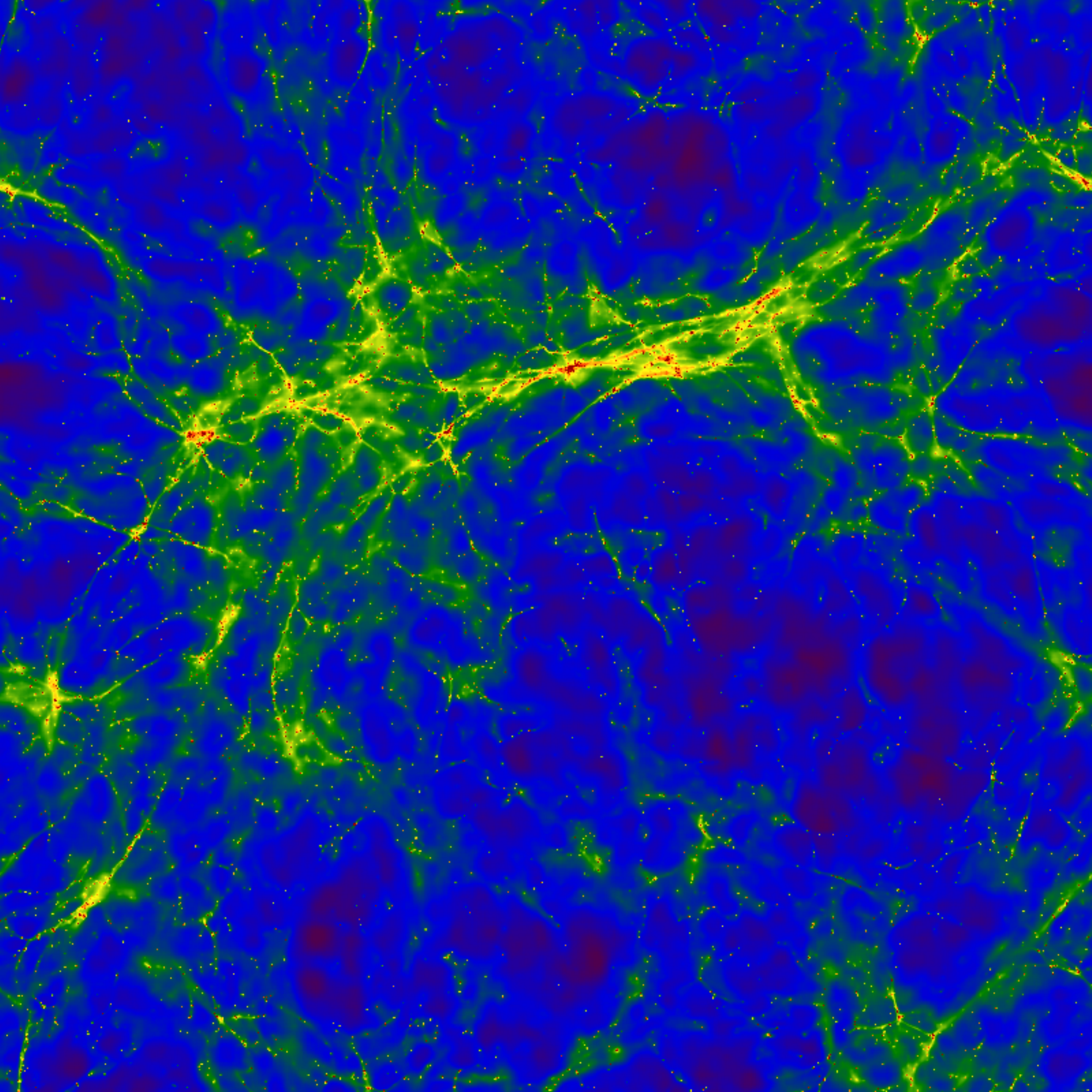}}
		\subfigure{\hfill\includegraphics[width=0.5\columnwidth]{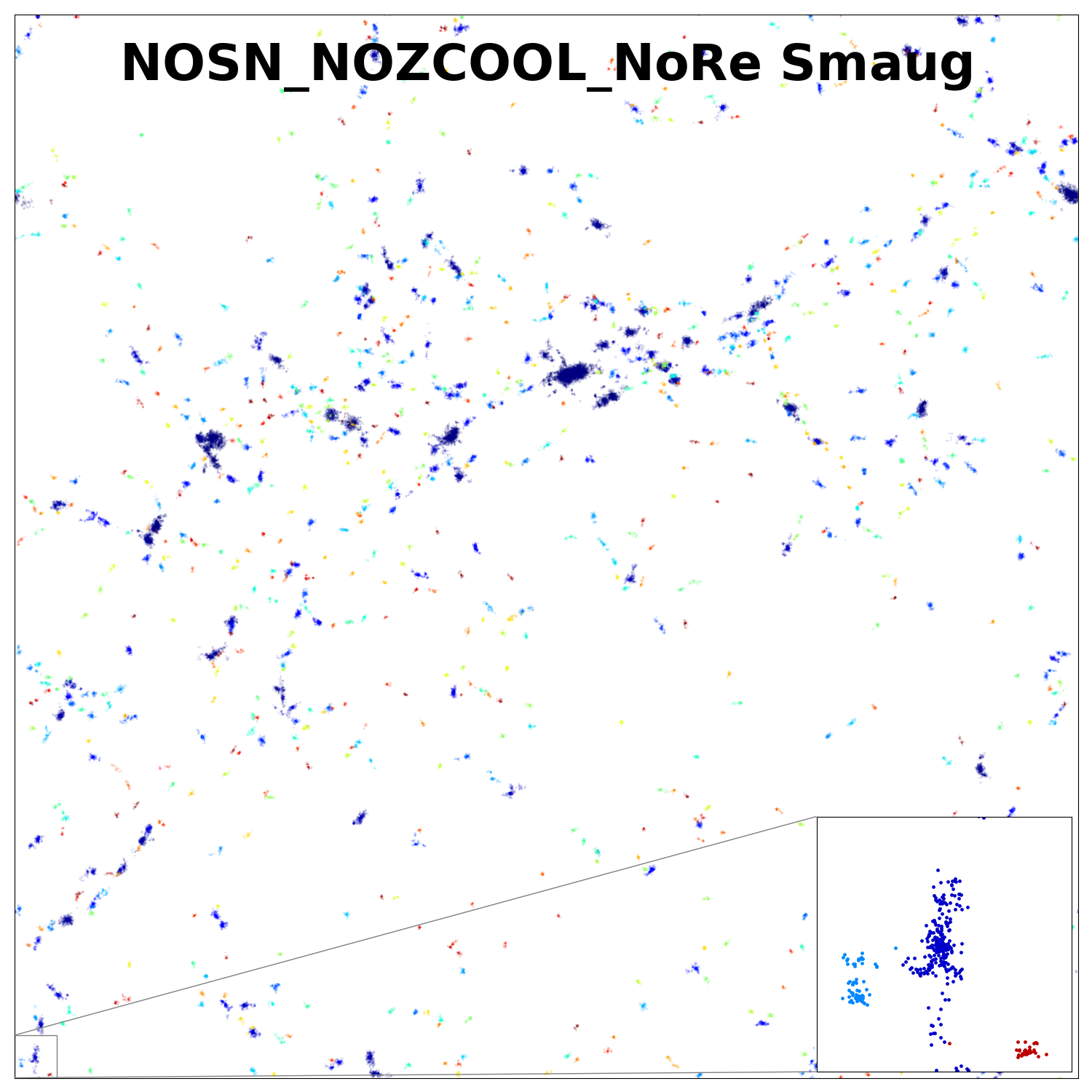}}\\
		\subfigure{\includegraphics[width=0.5\columnwidth]{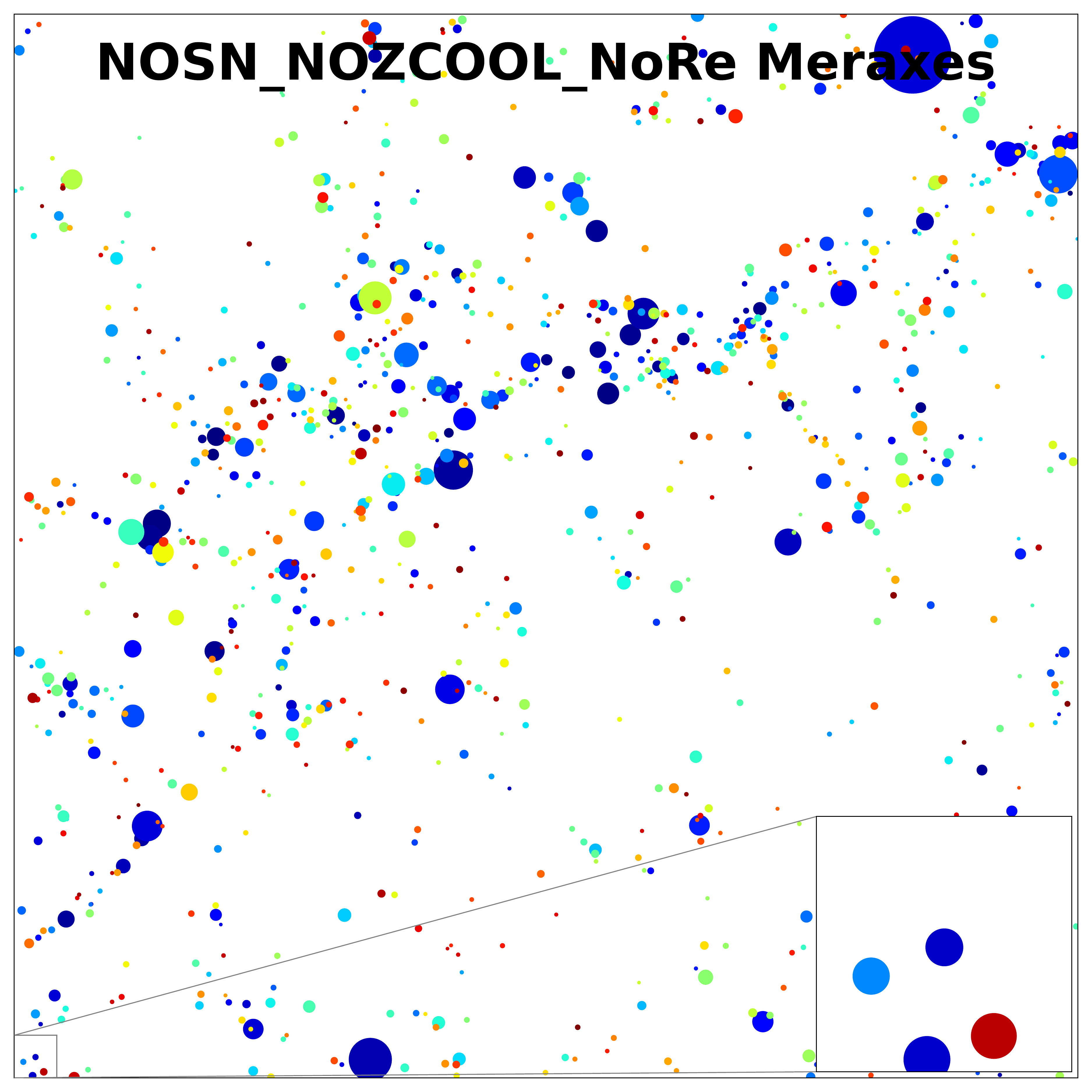}}
		\subfigure{\hfill\includegraphics[width=0.5\columnwidth]{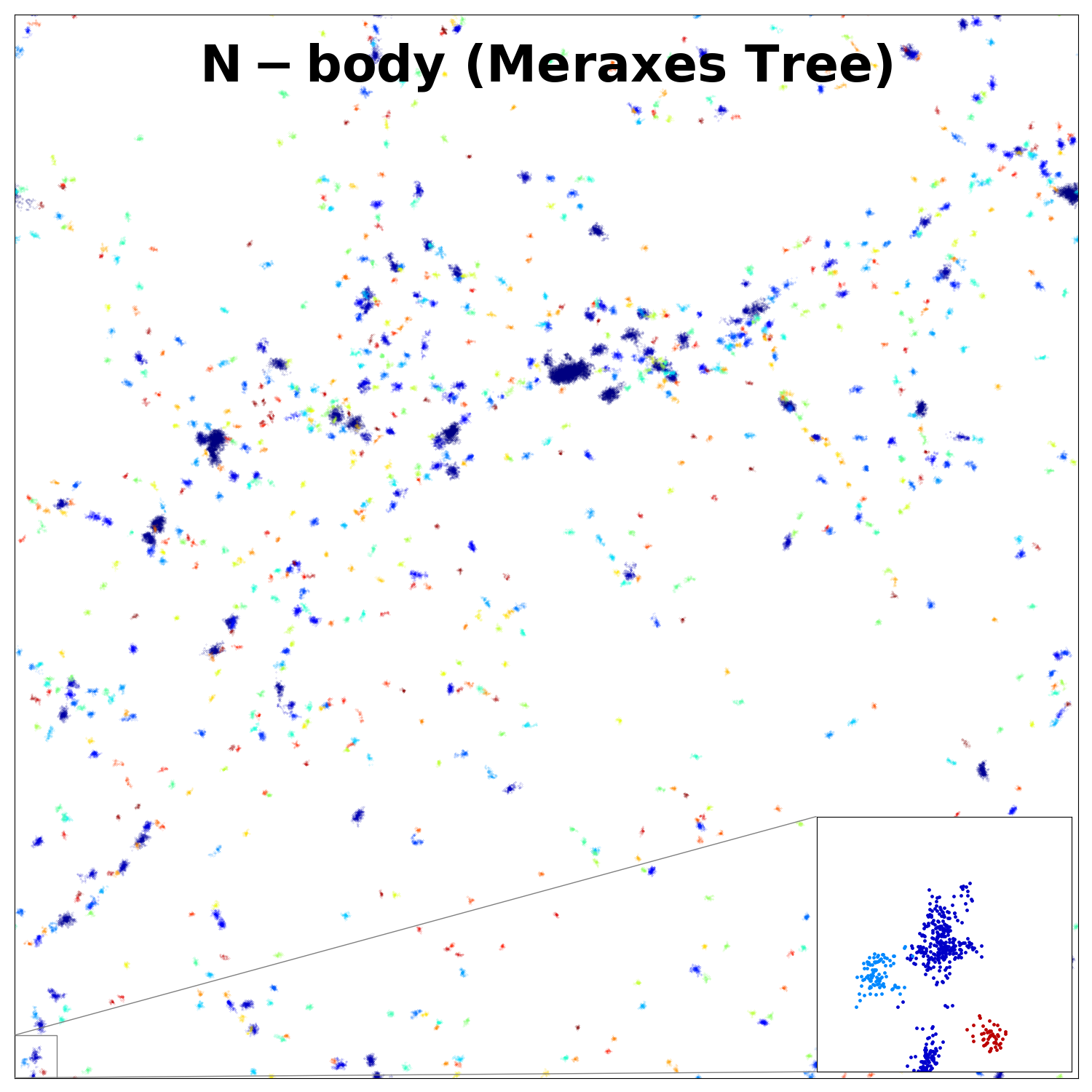}}
	\end{minipage}
	\caption{\label{fig:indicator}The top left panel shows the gas density distribution of the entire \textit{NOSN\_NOZCOOL\_NoRe} {\smaug} simulation volume at $z=5$. The remaining plots show an example of matched galaxies/haloes between three simulations (i.e. \textit{NOSN\_NOZCOOL\_NoRe} {\meraxes}, bottom left; the \textit{N}-body simulation, bottom right; and \textit{NOSN\_NOZCOOL\_NoRe} {\smaug}, top right). The bottom left shows the distribution of the first matched 1000 galaxies in {\meraxes} with circle size representing stellar mass. Varied colours indicate different objects for better visualization. The bottom right shows the distribution of the dark matter particles within the host haloes of the aforementioned 1000 galaxies in the \textit{N}-body simulation. The top right panel shows how the particles of those 1000 haloes in the bottom right panel are distributed in the \textit{NOSN\_NOZCOOL\_NoRe} {\smaug} simulation. The zoom-in boxes on the bottom right indicate four matching candidates.}
\end{figure*}

In this work, we make direct comparisons of the semi-analytic and hydrodynamic galaxy properties calculated by {\meraxes} and {\smaug}, respectively. We connect galaxies produced by {\meraxes} to their host haloes in the \textit{N}-body simulation, and then match galaxies in {\meraxes} with the corresponding hydrodynamic simulation. We describe the method in detail for matching galaxies between {\meraxes} and {\smaug} in this section.

\subsection{Running \textsc{Meraxes} on dark matter halo merger trees from {\smaug}}\label{sec:gbptrees}
The same algorithm of matching haloes between snapshots is adopted to match between simulations. Therefore, we first discuss the construction of halo merger trees in the DRAGONS framework (\citealt{Poole2016,poole2017mnras.472.3659p}).

\subsubsection{Dark matter halo merger trees}\label{sec:dark matter halo merger trees}
Dark matter (and baryon) particles evolve (co-evolve) in {\smaug}, and friend-of-friend (or {\sc fof}) groups and subgroups (or haloes), to which particles belong are determined using a standard {\sc fof} halo finder and subgroup finder \citep[{\sc subfind},][]{springel2008aquarius} in post-processing. More specifically, the halo finder identifies {\sc fof} groups using a standard linking length of 0.2 and then treats each group as a sphere with centre located at the position of its most bounded particle (\textit{MBP}). Then by tracking halo particles in consecutive snapshots, each halo can be linked to its progenitors and form a merger tree. Each merger tree can be traced towards higher redshifts until no progenitors can be identified. This process is applied to all haloes at one snapshot, in order to horizontally construct the merger trees, which allows the SAM to evolve all galaxies at each snapshot and calculate reionization feedback self-consistently. We note that the {\smaug} trees provide 103 snapshots between $z=50$ and 5 with a time interval of ${\sim}11.3$ Myr. This high temporal resolution is considered a major improvement of the DRAGONS project (\citetalias{Mutch2016a}), which is able to resolve the dynamical time of the galactic disc at high redshift (see equation \ref{eq:sfr}).

\subsubsection{Matching haloes}\label{sec:matching}

We next give a brief overview of the {\sc gbptrees} algorithm\footnote{\url{https://github.com/gbpoole/gbpCode}} used to build dark matter halo merger trees, which is essential for understanding the matching process. We refer the interested reader to \citet{poole2017mnras.472.3659p} for more detailed descriptions and numerical tests. 

For each halo, {\sc subfind} sorts particles according to their kinetic and potential energies, with, in general, more bounded particles ranking higher. More specifically, we consider a halo, $\mathscr{H}^{\left(i\right)}$ at snapshot $i$, which consists of $n^{\left(i\right)}$ particles radially sorted in order of decreasing potential energies with ranks of $r^{\left(i\right)} = 1, 2, ..., n^{\left(i\right)}$. At snapshot $j\ne i$, the particles of $\mathscr{H}^{\left(i\right)}$ can be located within several haloes $\{\mathscr{H}^{\left(j\right)}_{k}; k=1,2,...,n_k\}$, each of which is considered as a matching candidate (i.e. progenitor or descendant) and has $n^{\left(ij\right)}_k$ shared particles with $\mathscr{H}^{\left(i\right)}$. For each candidate, a pseudo-radial moment is defined as
\begin{equation}
S_k^{\left(ij\right)}\left(m\right) = \sum_{l=1}^{n_k^{\left(ij\right)}}\left[{r^{\left(i\right)}_l}\right]^m,
\end{equation} 
where $r^{\left(i\right)}_l$ is the rank of particle $l$ of halo $\mathscr{H}^{\left(j\right)}_k$ in halo $\mathscr{H}^{\left(i\right)}$. Therefore\footnote{$\sum_{k}S_k^{\left(ij\right)}$ is smaller than $S_\mathrm{max}^{\left(ij\right)}$ if some of the particles are in unresolved haloes.}, $\sum_{k}S_k^{\left(ij\right)}\leq S_\mathrm{max}^{\left(ij\right)}\equiv1^m+2^m+...+{n^{\left(i\right)}}^m$. We define the goodness of a matching candidate as 
\begin{equation}
\begin{split}
&\Delta f_k^{\left(ij\right)} {\equiv} \left.f_k^{\left(ij\right)}\right\vert_{m{=}{-}1}{-}\left.f_k^{\left(ij\right)}\right\vert_{m{=}0},\\
&\mathrm{where}\\
&f_k^{\left(ij\right)}\left(m\right) = \dfrac{S_k^{\left(ij\right)}\left(m\right)}{S_\mathrm{max}^{\left(ij\right)}\left(m\right)}.
\end{split}
\end{equation}
$\mathscr{H}^{\left(i\right)}$ and $\mathscr{H}^{\left(j\right)}_k$ are considered as a good match when $\Delta f_k^{\left(ij\right)}>-0.2$ \citep{poole2017mnras.472.3659p} Amongst all good matches, the best match is identified by maximizing the statistic of $S_k^{\left(ij\right)}\left(m{=}{-}1\right)$ during the process of scanning for good matches forwards and backwards over 16 snapshot\footnote{When $i$ is close to 0 or the total number of snapshots, $|j-i|<16$, leading to a worse matching result. This usually affects the performance of the halo merger trees when the simulation reaches the end.}, $|j-i|=1,2,...,16$. 

A consequence of this algorithm is that matching is performed by tracking halo cores instead of the majority of halo particles. This central weighting algorithm allows us to build more realistic halo merger trees, distributed on which galaxies form and evolve within the central regions of their host haloes. However, pathologies can also arise in this algorithm, which require special treatments. We briefly review two consequences of central weighting while more dedicated tests and analysis are presented in \citet{poole2017mnras.472.3659p}.

\begin{enumerate}
	\item When a halo orbits around a more massive object for a period of time without a subsequent merger event, it might be identified as a fraction of the larger system by {\sc subfind}, leading to a temporary merger event and causing numerical confusions. Under this circumstance and when the two haloes split, the descendant of the temporarily merged system is determined by the dominant object in the central regions before the separation. However, a side effect arises as well that the emerging halo is identified as a newly formed system without any progenitors. Note that losing progenitors in the merger trees leads to a reset of the hosted galaxy and thus discards its entire evolutionary history, which is catastrophic for semi-analytic galaxy formation models. The same impact also arises when the descendant cannot be identified due to tidal disruptions or numerical noise. Therefore,{\sc gbpTrees} is optimized to capture these and link the progenitor or locate the descendant with matching dedicated over consecutive and multiple snapshots.
	
	\item When a central halo, which is usually considered as the most massive halo in an {\sc fof} group and dominates the gravitational potential of the entire system, is interacting with its satellites, particles are frequently exchanged between central and satellites. This can potentially lead to numerical confusion that the dominant halo switches between different substructures of the {\sc fof} group while their cores remain nearly intact. In this case and because matching performed by following the core restricts galaxies in the centres of their host haloes, pathologies might arise that satellite galaxies receive unrealistic increments of mass on their hosts, leading to massive infall from the IGM and dramatic star formation when cooling is efficient.
\end{enumerate}	

\subsection{Match between \textsc{Meraxes} and {\smaug}}\label{sec:match between meraxes and smaug}

\begin{figure}
	\centering
	\includegraphics[width=\columnwidth]{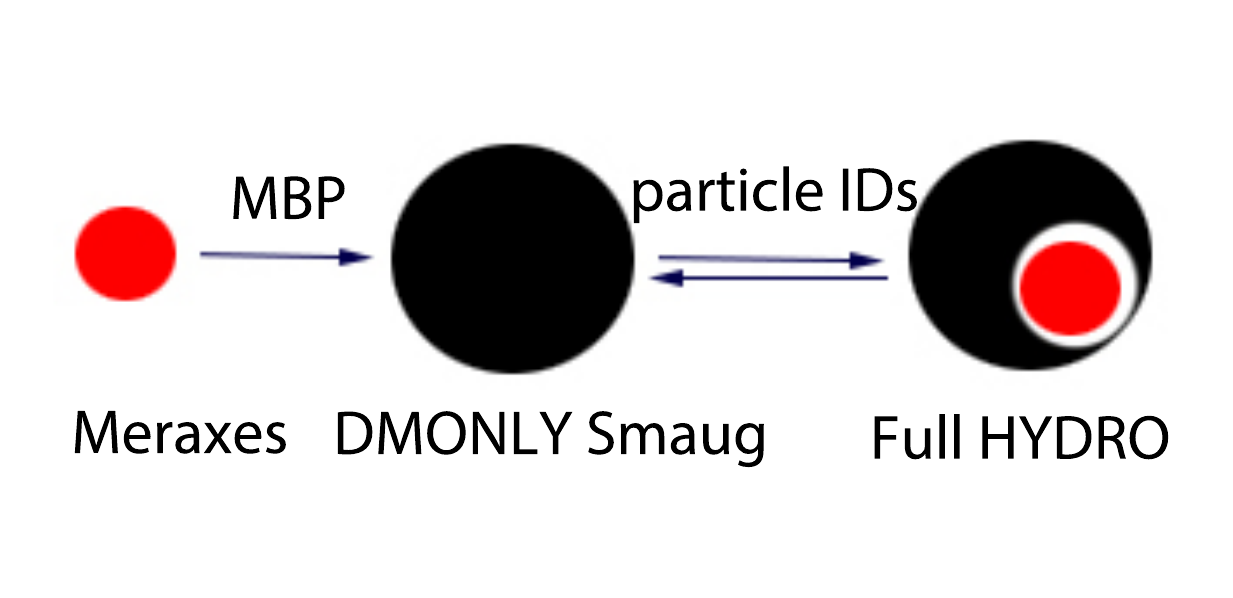}
	\caption{\label{fig:match} Illustration of matching between galaxies in {\meraxes} and full-hydrodynamic simulation {\smaug} outputs. The red and black circles indicate galaxies and their host haloes, respectively. We link galaxies in {\meraxes} with their host haloes in the \textit{N}-body simulation through the most bonded particle, \textit{MBP}. Then haloes between \textit{N}-body and full-hydrodynamic simulations are matched using dark matter particle IDs.}
\end{figure}

{\meraxes} reports the \textit{MBP} of the host halo for each galaxy, which bridges SAM galaxies and dark matter haloes in \textit{DMONLY}. In order to further connect haloes in the collisionless \textit{N}-body simulation with simulations including baryonic physics (i.e. \textit{NOSN\_NOZCOOL\_NoRe}), we match haloes between simulations using the same matching strategy described in Appendix \ref{sec:gbptrees}. Instead of matching 33 snapshots in one simulation, we match haloes from two simulations (e.g. \textit{NOSN\_NOZCOOL\_NoRe} and \textit{DMONLY}) using their dark matter particles at each snapshot. This is achievable because all the {\smaug} simulations including \textit{DMONLY} start from the identical cosmological initial conditions and their dark matter particles possess the same IDs over different simulations. However, with information from only two snapshots, mismatches and pathologies occur in particular when reaching the resolution limit. Therefore, final matched sample is limited to haloes which are matched bidirectionally. Fig. \ref{fig:match} illustrates how galaxies are finally matched between {\meraxes} and {\smaug} with this two-step manner that \textit{MBP}s connect galaxies from {\meraxes} to their host haloes which are matched with {\smaug} galaxies through dark matter particles. In this work, only central galaxies with virial masses exceeding the resolution threshold of $10^{7.5}\mathrm{M}_\odot$ in the SAM results are considered, and in order to minimize the impact from central-satellites switching mentioned in Appendix \ref{sec:gbptrees}, satellite galaxies identified in the SAM are excluded from the final sample as well.

The right-hand panels of Fig. \ref{fig:indicator} project the spatial distributions of dark matter particles in the \textit{NOSN\_NOZCOOL\_NoRe} and \textit{N}-body simulations from {\smaug} on to the xy-plane, which illustrates how well the matching procedure works. In Fig. \ref{fig:indicator}, the bottom left panel shows the distribution of the first matched 1000 galaxies\footnote{The position of galaxy in {\meraxes} is inherited from its host halo in the \textit{N}-body simulation.} in {\meraxes} (also without supernova feedback and metal cooling) with circle size representing stellar mass and different colours for better visualization. The bottom right panel shows the distribution of dark matter particles of the corresponding host haloes in the \textit{N}-body simulation while the top right panel shows how these particles are distributed in the \textit{NOSN\_NOZCOOL\_NoRe} {\smaug} simulation. The bottom right of each panel zooms into four objects for better illustration. In the \textit{NOSN\_NOZCOOL\_NoRe} {\smaug} simulation, there are 195562 galaxies, in which 129043 galaxies ($\sim 66$ per cent) are matched with {\meraxes}.
         
\section{Hot and cold non-star-forming gas}\label{app:sec:hot}

\begin{figure}
	\begin{minipage}{\columnwidth}
		\centering
		\includegraphics[width=\textwidth]{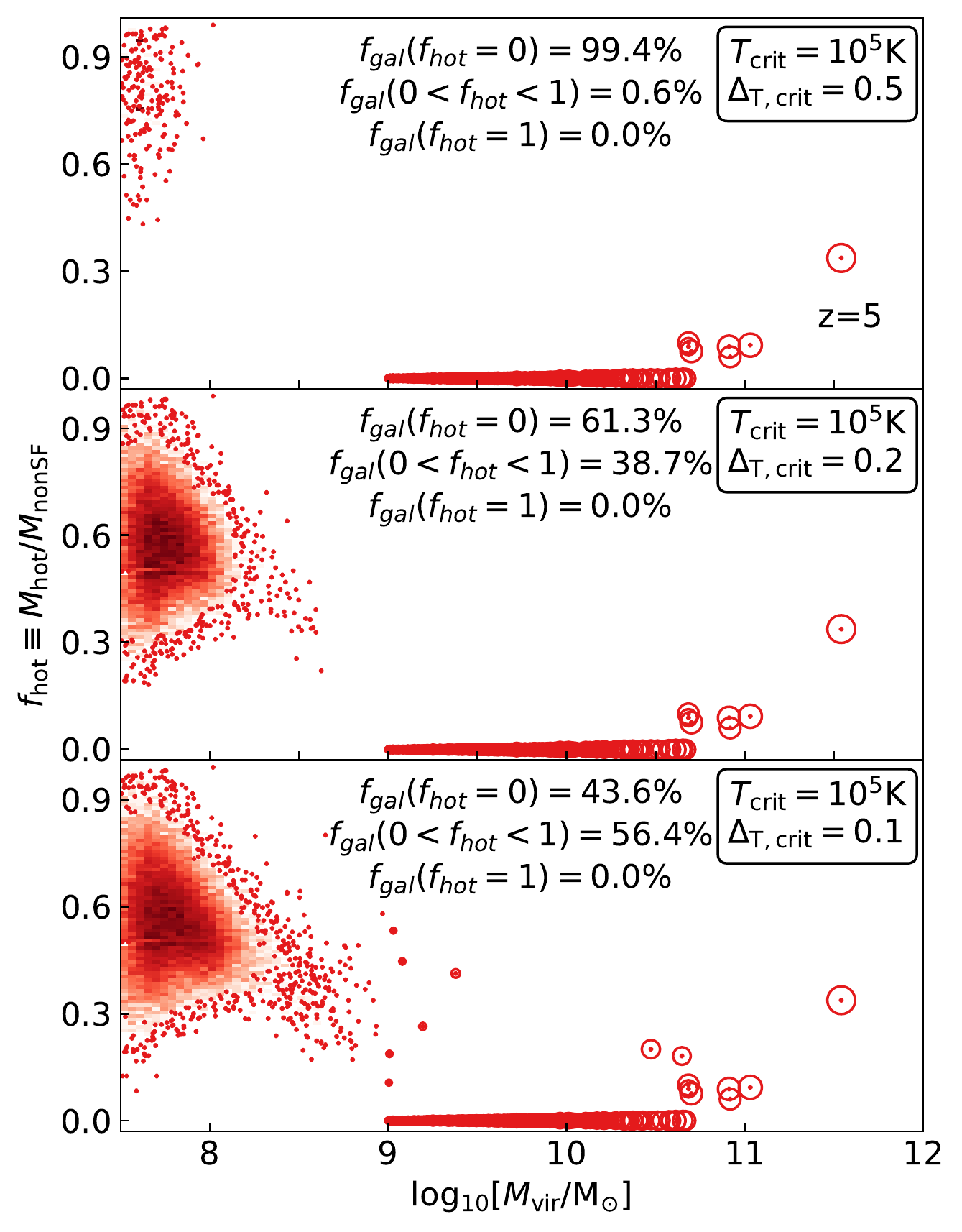}		
	\end{minipage}
	\caption{\label{app:fig:fhot_DeltaTcrit}The fraction of hot gas in the non-star-forming gas ($f_\mathrm{hot}\equiv M_\mathrm{hot}/M_\mathrm{nonSF}$) from the \textit{NOSN\_NOZCOOL\_NoRe} {\smaug} simulation at $z=5$ as a function of the halo mass ($M_\mathrm{vir}$). From top to bottom, $\Delta_\mathrm{T, cirt}$ is 0.5, 0.2 and 0.1. Galaxies with $f_\mathrm{hot}\ne0$ are indicated with red contours (logarithm scale) and dots at low-dense regions. Galaxies with $M_\mathrm{vir}>10^9\mathrm{M}_\odot$ are emphasized with red circles with circle size representing the stellar mass ($M_*$). The fraction of galaxies with $f_\mathrm{hot}=0$, $0<f_\mathrm{hot}<1$ and $f_\mathrm{hot}=1$ are shown in each subpanel. }
\end{figure}

In this work, we use a machine learning algorithm to spilt non-star-forming gas particles into hot and cold components. The two groups are considered as one when the offset between their median temperatures is small (i.e. less than $\Delta_\mathrm{T, cirt}$ in logarithm). Fig. \ref{app:fig:fhot_DeltaTcrit} shows the fraction of hot gas in the non-star-forming gas ($f_\mathrm{hot}\equiv M_\mathrm{hot}/M_\mathrm{nonSF}$) at $z=5$ with different choices of $\Delta_\mathrm{T, cirt}$ and we see that a smaller $\Delta_\mathrm{T, cirt}$ introduces more galaxies with $f_\mathrm{hot}>0$. However, this only affects galaxies with inefficient cooling (i.e. below the atomic cooling thresholds) while $f_\mathrm{hot}$ of more massive galaxies remain zero or small. On the other hand, when non-star-forming gas particles of a galaxy are considered as one group, which is common for galaxies with efficient cooling, whether they are determined as hot or cold (i.e. $f_\mathrm{hot}=1$ or 0) is based on their median temperature. We vary the minimum temperature of particles being identified as hot ($T_\mathrm{crit}$) from $10^5\mathrm{K}$ to $10^4\mathrm{K}$ and we find that $f_\mathrm{hot}$ is insensitive to $T_\mathrm{crit}$. Based on these, we note that the choice of $\Delta_\mathrm{T, cirt}$ and $T_\mathrm{crit}$ does not have a significant impact to our conclusion that high-redshift dwarf galaxies accrete gas in cold mode (see Section \ref{sec:gas_infall1}).

\section{Semi-analytic properties}\label{app:sam_properties}

\begin{figure}
	\begin{minipage}{\columnwidth}
		\centering
		\includegraphics[width=\textwidth]{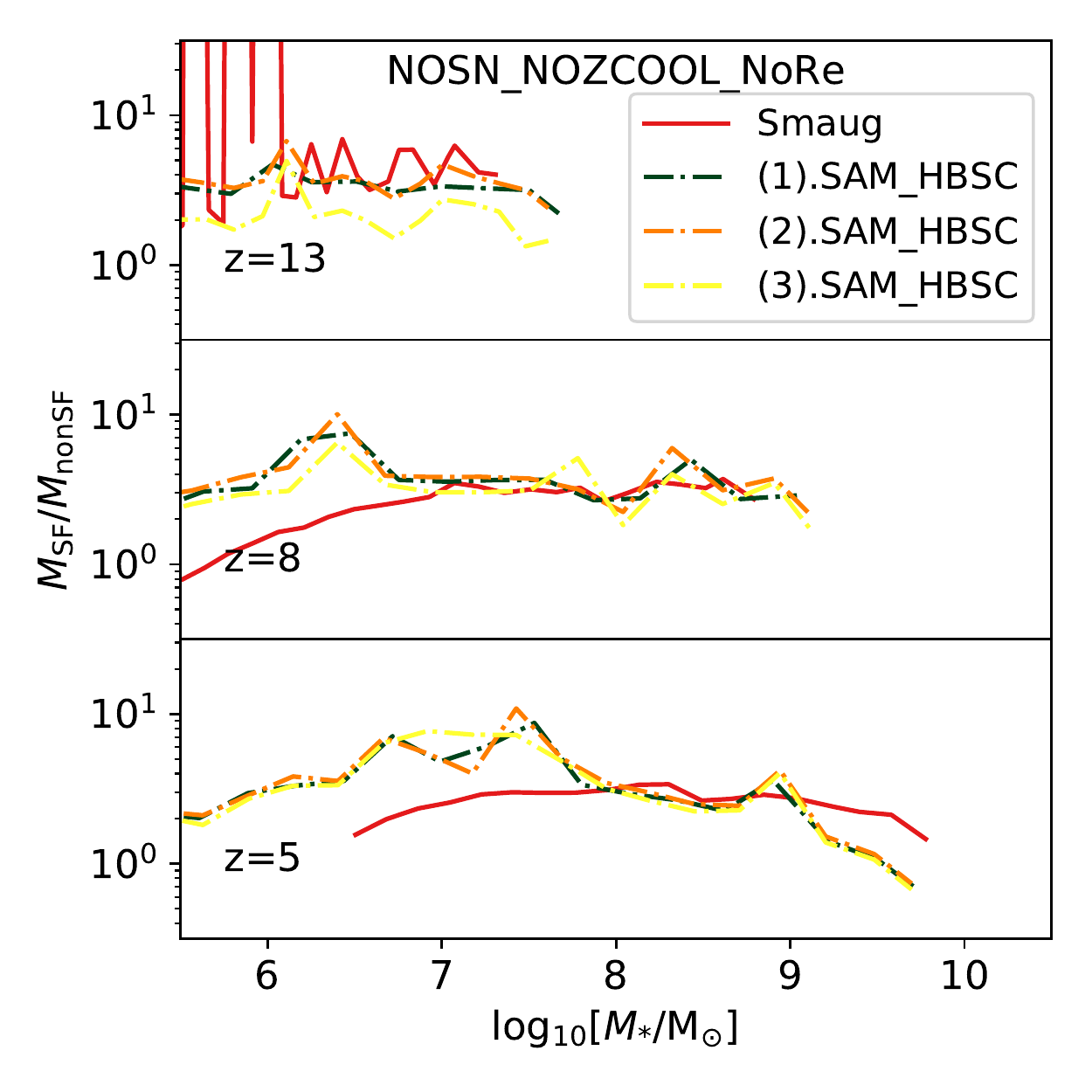}%\vspace*{-3mm}
	\end{minipage}
	\caption{\label{app:fig:sf_nosn_nozcool_nore}The mass ratio of star-formation gas to non-star-forming gas as a function of stellar mass at $z=13-5$ from {\meraxes} and {\smaug} \textit{NOSN\_NOZCOOL\_NoRe} results. Three {\meraxes} results with redshift-dependent star formation efficiency and cooling rate (\textit{SAM\_HBSC}) are shown with green dashed for calculations performed at different moments.}
\end{figure}

A numerical experiment evolves a galaxy by calculating physics processes in a predefined order. For instance, {\meraxes} in sequence calculates reionization, baryonic infall, cooling, star formation and the impact of supernova feedback. Thanks to the 
exquisite time step of calculation (not output) of hydrodynamic simulations, their outputs can be considered as instantaneous galaxy properties at a given redshift. However, this is not the case for {\color{black}the SAM used in this work}, which records properties that either have (e.g. stellar mass and gas mass) or have not been processed (e.g. host halo properties) or are average values (e.g. SFR) during one time step. For instance, in the right panel of Fig. \ref{fig:sf_nosn_nozcool_nore}, we show $M_\mathrm{SF}/M_\mathrm{nonSF}$, which is the mass ratio of cold to hot gas in the SAM. Three moments of calculation can be adopted:
\begin{enumerate}
	\item[(1)] after star formation, when both hot gas and cold gas have been consumed. In this case, gas mass is underestimated due to the lack of replenishment;
	\item[(2)] before star formation and after cooling, when hot gas has transitioned to cold gas. In this case, stellar mass and hot gas are underestimated while cold gas is overestimated;
	\item[(3)] before cooling and after baryon infall, when hot gas has been refurbished but cold gas still represents the remaining gas reservoir from the previous snapshot or ${\sim}11$Myr ago. In this case, stellar mass and cold gas are underestimated while hot gas is overestimated.
\end{enumerate}
We note that case (1) is adopted in the main context, and we show the difference in Fig. \ref{app:fig:sf_nosn_nozcool_nore}. We see that because of the high cadence (${\sim}11$Myr) of the SAM, the choice of calculation does not have a significant impact to our results.

\bsp
\label{lastpage}
\end{document}